\documentclass[structabstract]{aa}  
\usepackage{graphicx}
\usepackage{txfonts}
\usepackage{lscape}
\begin{document}
\newcommand{\as}[2]{$#1''\,\hspace{-1.7mm}.\hspace{.1mm}#2$}
\newcommand{\am}[2]{$#1'\,\hspace{-1.7mm}.\hspace{.0mm}#2$}
\def\approxlt{\lower.2em\hbox{$\buildrel < \over \sim$}}
\def\approxgt{\lower.2em\hbox{$\buildrel > \over \sim$}}
\newcommand{\dgr}{\mbox{$^\circ$}}   
\newcommand{\grd}[2]{\mbox{#1\fdg #2}}
\newcommand{\gsim}{\stackrel{>}{_{\sim}}}
\newcommand{\HI}{\mbox{H\,{\sc i}}}
\newcommand{\HII}{\mbox{H\,{\sc ii}}}
\newcommand{\HIbf}{\mbox{H\hspace{0.155 em}{\footnotesize \bf I}}}
\newcommand{\HIit}{\mbox{H\hspace{0.155 em}{\footnotesize \it I}}}
\newcommand{\HIsl}{\mbox{H\hspace{0.155 em}{\footnotesize \sl I}}}
\newcommand{\IHI}{\mbox{${I}_{HI}$}}
\newcommand{\Jykms}{\mbox{Jy~km~s$^{-1}$}}
\newcommand{\kms}{\mbox{km\,s$^{-1}$}}
\newcommand{\kmsMpc}{\mbox{ km\,s$^{-1}$\,Mpc$^{-1}$}}
\def\lir{{\hbox {$L_{IR}$}}}
\def\lco{{\hbox {$L_{CO}$}}}
\def \ls{\hbox{$L_{\odot}$}}
\newcommand{\LB}{\mbox{$L_{B}$}}
\newcommand{\LBnul}{\mbox{$L_{B}^0$}}
\newcommand{\LBsun}{\mbox{$L_{\odot,B}$}}
\newcommand{\lsim}{\stackrel{<}{_{\sim}}}
\newcommand{\LsunK}{\mbox{$L_{\odot, K}$}}
\newcommand{\LsunB}{\mbox{$L_{\odot, B}$}}
\newcommand{\LsunMsun}{\mbox{$L_{\odot}$/${M}_{\odot}$}}
\newcommand{\LK}{\mbox{$L_K$}}
\newcommand{\LKLB}{\mbox{$L_K$/$L_B$}}
\newcommand{\LKLBnul}{\mbox{$L_K$/$L_{B}^0$}}
\newcommand{\LKLsun}{\mbox{$L_{K}$/$L_{\odot,Bol}$}}
\newcommand{\masq}{\mbox{mag~arcsec$^{-2}$}}
\newcommand{\MHI}{\mbox{${M}_{HI}$}}
\newcommand{\MHILB}{\mbox{$M_{HI}/L_B$}}
\newcommand{\MHILBfr}{\mbox{$\frac{{M}_{HI}}{L_{B}}$}}
\newcommand{\MHILK}{\mbox{$M_{HI}/L_K$}}
\newcommand{\KMS}{\mbox{$\frac{km}{s}$}}
\newcommand{\JYKMS}{\mbox{$\frac{Jy km}{s}$}}
\newcommand{\MHILKfr}{\mbox{$\frac{{M}_{HI}}{L_{K}}$}}
\def \ms{\hbox{$M_{\odot}$}}
\newcommand{\Msun}{\mbox{${M}_\odot$}}
\newcommand{\MsunLsun}{\mbox{${M}_{\odot}$/$L_{\odot,Bol}$}}
\newcommand{\MsunLBsun}{\mbox{${M}_{\odot}$/$L_{\odot,B}$}}
\newcommand{\MsunLKsun}{\mbox{${M}_{\odot}$/$L_{\odot,K}$}}
\newcommand{\MT}{\mbox{${M}_{ T}$}}
\newcommand{\MTLBnul}{\mbox{${M}_{T}$/$L_{B}^0$}}
\newcommand{\MTLBsun}{\mbox{${M}_{T}$/$L_{\odot,B}$}}
\newcommand{\nan}{Nan\c{c}ay}
\newcommand{\tis}[2]{$#1^{s}\,\hspace{-1.7mm}.\hspace{.1mm}#2$}
\newcommand{\Vcor}{\mbox{${V}_{0}$}}
\newcommand{\vhel}{\mbox{$V_{hel}$}}
\newcommand{\VHI}{\mbox{${V}_{HI}$}}
\newcommand{\vrot}{\mbox{$v_{rot}$}}
\def\la{\mathrel{\hbox{\rlap{\hbox{\lower4pt\hbox{$\sim$}}}\hbox{$<$}}}}
\def\ga{\mathrel{\hbox{\rlap{\hbox{\lower4pt\hbox{$\sim$}}}\hbox{$>$}}}}

 \title{H{\Large \bf I} line observations of 2MASS galaxies in the Zone 
        of Avoidance}

   \author{W. van Driel\inst{1,2},
          S.E. Schneider\inst{3},
          R.C. Kraan-Korteweg\inst{4}, 
          \and
          D. Monnier Ragaigne\inst{5}
          } 

  \offprints{W. van Driel}

  \institute{GEPI, Observatoire de Paris, CNRS, Universit\'e Paris Diderot, 
             5 place Jules Janssen, 92190 Meudon, France  \\
             \email{wim.vandriel@obspm.fr}
  	   \and  
  	         Station de Radioastronomie de \nan, Observatoire de Paris,
  	         CNRS/INSU, 18330 \nan, France 
  	   \and
             University of Massachusetts, Astronomy Program, 536 LGRC, Amherst, 
             MA 01003, U.S.A.  \\
             \email{schneide@messier.astro.umass.edu}
       \and
            Department of Astronomy, University of Cape Town, 
			      Private Bag X3, Rondebosch 7701, South Africa  \\
            \email{kraan@ast.uct.ac.za}
       \and
            Laboratoire de l'Acc\'el\'erateur Lin\'eaire, Universit\'e Paris-Sud 
            B\^atiment 200, BP 34, 91898 Orsay Cedex, France   \\
            \email{monnier@lal.in2p3.fr}
       }

\date{Received 23 March 2009 / Accepted 7 May 2009}

\abstract
{}
{A pilot survey has been made to obtain 21cm \HI\ emission line profiles for 
197 objects in the Zone of Avoidance (ZoA) that were classified as galaxies in
the 2MASS all-sky near-infrared Extended Source Catalog (2MASX), as well as
a further 16 2MASS pre-release Working Database sources that did not make it 
into 2MASX.}
{116 of the 2MASX sources and the 16 Working Database sources were observed 
using the \nan\ radio telescope, usually in the 325 to 11,825 \kms\ range, 
and the other 81 2MASX sources were observed with the Arecibo radio telescope in the 
--500 to 11,000 \kms\ range, and for 9 also in the 9,500 to 21,000 \kms\ range. }
{Global \HI\ line parameters are presented for the 22 and 29 2MASX objects
that were detected at \nan\ and Arecibo, respectively, as well as
upper limits for the undetected 2MASX objects. Another galaxy (ESO 371-27) 
was detected in the \nan\ beam centred on an undetected target, ESO 371-26.
\nan\ data on 12 sources could not be used due to high rms noise levels, most 
likely caused by strong nearby continuum sources. None of the 16 Working Database 
sources were detected at \nan.
Whereas object 2MASX J08170147-3410277 appears to be a very massive galaxy with 
an \HI\ mass of $4.6 \cdot 10^{10}$ \Msun\ and an inclination-corrected rotation 
velocity of 314 \kms, it is clear that only radio synthesis \HI\ imaging observations 
will allow a firm conclusion on this.
}
{Overall, the global properties of the detected galaxies match those of other ZoA \HI\ 
surveys. Although the detections are as yet too sparse to give further insight into 
suspected or unknown large-scale structures in the ZoA, they already indicate that 
an extension of the present pilot survey is bound to quantify filaments, clusters, 
and voids behind this part of the Milky Way. 
 It is shown that the number of candidate 2MASS-selected ZoA galaxies to be observed in 
\HI\ could have been reduced by about 15\% through examination of composite 
near-infrared images and the application of extinction-corrected near-infrared colour limits. 
{\rm Present results confirm that the Galactic extinction values from 
Schlegel et al. (1998) are valid for latitudes $|b| \ga 5\dgr$, but increasingly 
less so for lower latitudes.} 
} 
   

 \authorrunning{van Driel et al.}
 \titlerunning{\HI\ line observations of 2MASS galaxies in the ZoA}
 
 \maketitle

%

\section{Introduction} 
Because of dust extinction at low Galactic latitudes, resulting in the
so-called Zone of Avoidance (ZoA), redshift surveys have generally
concentrated on regions farther than 10$^{\circ}$ from the Galactic
plane (e.g. Kraan-Korteweg \& Lahav 2000), except for systematic
\HI\ surveys (e.g. HIPASS, Meyer et al. 2004) and systematic \HI\ ZoA
at $|b| \le 5\dgr$ (Henning et al. 2000a,b; Donley et al. 2005; Kraan-Korteweg 2005). 
This has left a gap in our knowledge of local large-scale structure 
over a large part of the sky -- see, e.g., Fig. 10. The existence and membership
of nearby groups of galaxies is still highly uncertain and even dwarfs
within the Local Group keep being detected at low latitudes. We have collected 
21cm \HI\ line data of near-infrared selected galaxies with unknown redshifts 
in the ZoA ($|b| \le 10$\dgr), mostly focusing on objects with relatively large angular size that 
are likely to be nearby, to reduce the redshift ZoA and complement the various 
existing all-sky (redshift) surveys.

The interstellar extinction in the $K_s$-band (2.2$\mu$m) is 12 times
smaller than in the $B$-band and 5.5 times smaller than in the $I$-band (Cardelli et al. 1989). 
We therefore decided to select probable ZoA galaxies from the -- at that time (in 2000-2002) 
-- emerging systematic 2-Micron All Sky Survey (2MASS; Skrutskie et al. 2006). Many of these
ZoA galaxies are invisible on the Palomar Sky Survey (DSS) images and
even in the $I$-band have several magnitudes of extinction, whereas the extinction in the 
$K_s$-band remains relatively modest in most cases. Follow-up \HI-observations of these 
heavily absorbed galaxies are then the most efficient -- if not only -- tool to obtain 
a distance estimate and map large-scale structures across the ZoA.

Nevertheless, identification of galaxies at very low Galactic latitudes remains difficult 
even using 2MASS, not so much because of dust extinction, but due to stellar crowding 
close to the Galactic plane.
In particular, the wider Galactic Bulge area makes the automatic identification of extended 
sources increasingly uncertain, or even impossible when star densities of log$N=4.00$/(deg)$^2$ 
are reached for stars with $K \le 14\fm0$, leading to the so-called NIR ZoA (see
Fig.~9 in Kraan-Korteweg 2005).

2MASS used two identical telescopes in the north and the south to observe the entire sky at 
$J$, $H$, and $K_s$, providing an opportunity to select a much more uniform sample of galaxies 
than has been possible previously. Note that although we initially used the 2000 and 2002 versions of 
the working database for our source selection, all data presented here are updated to the 
final release catalogue values. Figure~1 shows an all-sky map of 2MASS-selected bright
($K_s$$\le$11\fm4), extended sources in Galactic coordinates (centred on the Galactic anti-centre direction), 
demonstrating that we can select sources quite uniformly deep into the ZoA. The heavy 
curves in the figure show declinations of --40$^{\circ}$, 0$^{\circ}$, and +35$^{\circ}$ 
which correspond to the declination limits of the radio telescopes (\nan\ and Arecibo; 
see Section~2 for further details) used in this pilot project of observing 2MASS-selected 
ZoA galaxies in \HI.  

2MASS has a 95\% completeness level of $K_s<13\fm5$ at high Galactic latitudes (Jarrett et al. 2000), 
and has little dust extinction at low latitudes. However, close to the Galactic plane, the 
numerous faint (red) foreground stars effectively create a high sky-brightness noise level, and 
the fainter (low surface brightness) galaxies are extremely difficult to detect because of confusion 
noise. We therefore preferentially selected galaxies with $K_s<12\fm0$ where the 2MASS galaxy 
sample still remains essentially complete (cf. Huchra et al. 2005). A $K_s=12\fm0$ completeness 
limits corresponds to an optical $B$-band magnitude limit of $16\fm0$ to $14\fm0$ respectively, 
for typical $B-K$ galaxy colours ranging from about $4\fm0$ for ellipticals to $2\fm0$ for spirals 
(e.g., Jarrett et al. 2003). 

In practice, a $K_s=12\fm0$ selection was possible for the \nan\ sample only, where 92\% meet 
this criterion, whereas finding enough sources not yet observed in \HI\ in the considerably 
smaller area observable at Arecibo resulted in selecting fainter sources, 93\% of which have 
$K_s>12\fm0$. 

It should be emphasized here, however, that a NIR selected ZoA galaxy sample may not necessarily 
be identical to objects selected at high Galactic latitudes. It likely is biased in the sense 
that such a survey favours galaxies with higher (redder) surface brightness, or shorter 
scale lengths, hence early type or bulge-dominated spirals, against low surface brightness 
bluer galaxies. This is difficult to determine directly because of uncertainties in estimates 
of Galactic extinction and stellar subtraction. In fact, this is already playing a role at high 
Galactic latitudes where 2MASX underestimates magnitudes of galaxies by 0\fm2 to 2\fm5 (for 
total magnitudes) due to the loss of other LSB features of extended dwarf galaxies, hence being 
increasingly incomplete for LSB galaxies, and fails to detect the lowest LSB galaxies entirely 
(Kirby et al. 2008; Andreon 2002). 

The near-infrared is less affected by a galaxy's star-formation history. A $K_s$-selected 
sample therefore provides a more accurate indication of the total stellar content of the galaxies, 
and is, for instance, less biased by galaxy interactions which can trigger star formation. 
Collecting redshifts and \HI\ parameters for these galaxies, in particular in the ZoA, also 
contributes to broader efforts to build complete all sky samples of bright 2MASS galaxies, 
such as the 2MASS Redshift Survey (2MRS; see Huchra et al. 2005).
Future studies can use the results from the present project to systematically identify a subset 
of 2MASS ZoA galaxies with good \HI\ characteristics, such as appropriate inclinations and 
strong \HI\ emission, that will allow the determination of peculiar velocities based on the 
NIR Tully-Fisher (TF) relation, and will complement ongoing efforts by Masters et al. (2008) 
for the whole sky 2MASS Tully Fisher Relation Survey (2MTF). In combination with the large-scale 
structure information provided by the 2MRS, such a survey will allow a significant improvement 
over dipole anisotropy derivations as well as reconstructed density and velocity fields from the 
2MASS Redshift Survey (Erdogdu et al. 2006a,b) for which the lack of data in the ZoA still remains 
one of the major contributors to the uncertainties in understanding the dynamics in the nearby 
Universe as well as the Local Group (Loeb \& Narayan 2008).

This paper is structured as follows: the selection of the sample of
2MASS ZoA galaxies observed by us in \HI\ with the \nan\ and Arecibo
radio telescopes is described in Section 2, the observations and the
data reduction are presented in Section 3, and the results in Section
4. A discussion of the results is given in Section 5 and the conclusions are presented in Section 6.

\section{Sample selection} 
\subsection{\nan\ sample} 
At \nan, we surveyed the ZoA within $\pm10^\circ$ of the Galactic plane in the declination 
zones of --39$^{\circ}$ to 0$^{\circ}$, and +35$^{\circ}$ to +73$^{\circ}$ (+73$^{\circ}$ being the 
northernmost limit of the $\pm$10$^{\circ}$ latitude band around the Galactic plane). The reason 
for the exclusion of the declination range 0\dgr\ to +35\dgr\ is obviously due to the Arecibo 
telescope being much more efficient in capturing galaxies there. This leads to a Galactic longitude 
coverage of selected objects in the ZoA of about $\ell$: -20\dgr\ (340\dgr) to 170\dgr\ and 
$\ell$: 210\dgr\ to 270\dgr. It should also be pointed out that the majority of the galaxies 
observed  in the northern Zone of Avoidance do not overlap too much with the systematic (blind) 
ZoA surveys pursued with the Parkes Multi-Beam Receiver in the south and its extension to the north 
($|b| \le 5$\dgr) as well as the extension to higher latitudes around the Galactic Bulge 
(see Kraan-Korteweg et al. 2008 for survey details). Indeed we have very few detections in 
common (see Section 5.5, Fig.~\ref{LSS}). This must be influenced by the fact that 2MASS has identified 
very few low latitude galaxies ($\la \pm 5$\dgr) in the southern sky, i.e. around the Galactic Bulge. 

\begin{figure} 
\centering
\includegraphics[width=9cm]{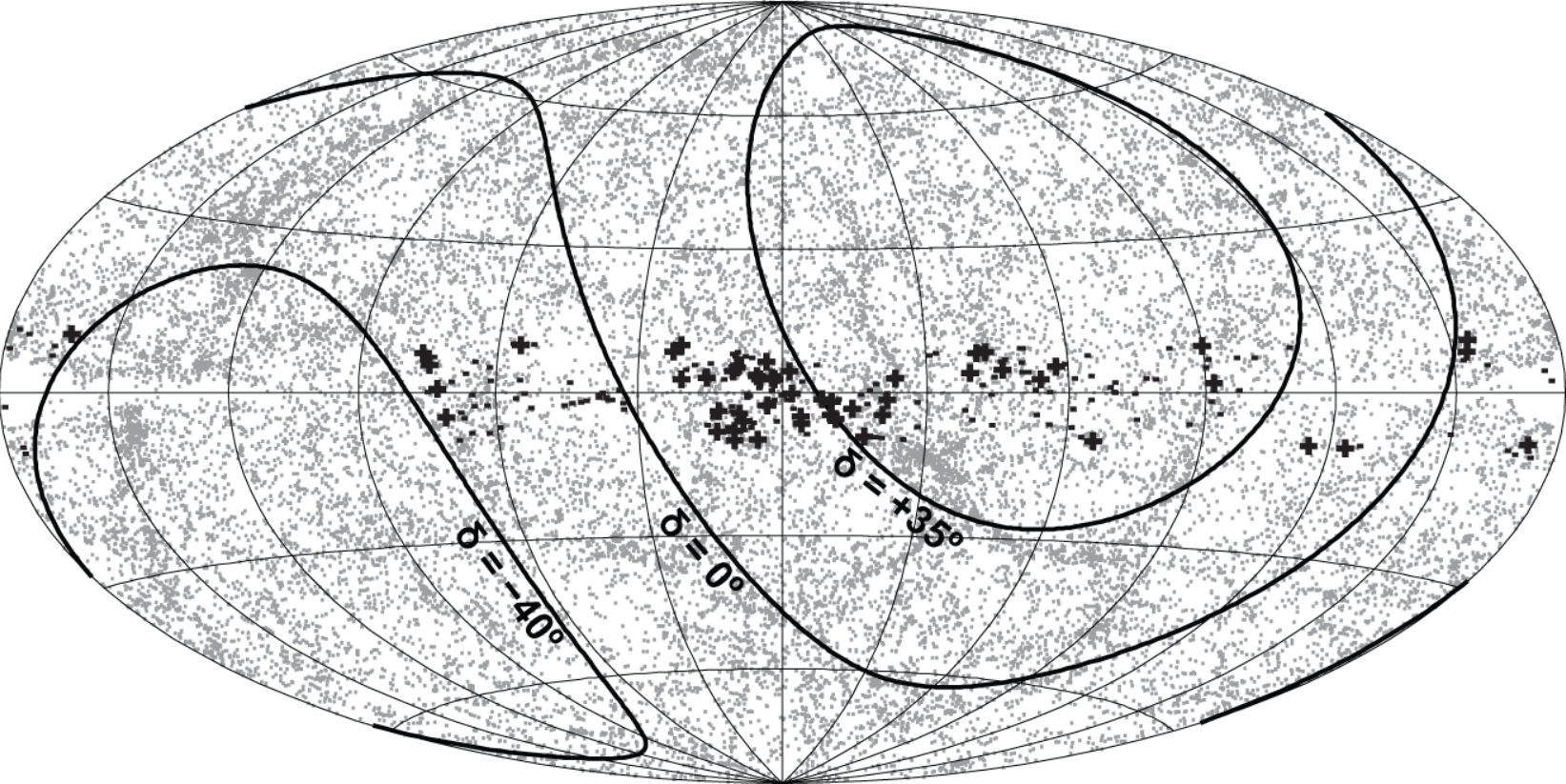}
\caption {All-sky plot of bright ($K_s$$\le$11\fm4) 2MASS extended sources. The galaxy positions 
(gray dots) are plotted in Galactic coordinates, centred on the Galactic anticentre. The black dots 
indicate galaxies observed in this project, and the + symbols indicate galaxies whose \HI\ signal 
was detected. }
\label{zoaHIfig}%
\end{figure}

Using a 2002 pre-release version of the 2MASS Working Database, for \nan\ we selected the 2MASS sources that 
were classified as galaxies with the largest angular radii ($>$\as{41}{8} at 20 \masq\ in the $K_s$-band) 
within the region to be covered at \nan\ ($\sim$3,600 deg$^2$). This led to over 8,000 galaxies with 
extinction-corrected $K_s<12\fm0$. 
We selected the strongest candidate galaxies from among these based on 
the classification scheme used to identify galaxies among the extended 
sources in the 2MASS catalogue (Jarrett et al. 2000). This automated 
classification scheme used training sets of known objects to assign a 
probability of a source being a galaxy or other extended source based on 
shape, symmetry, colour and other catalogued parameters.
We then chose the largest from among these sources, leaving 1448. 
We searched the NASA/IPAC Extragalactic Database (NED) for any matching sources within 3$'$ of the 
positions of these sources and found that 581 had been previously classified as galaxies, and 278 had 
published redshifts. Eliminating all sources with known redshifts, we selected the 300 largest remaining
sources, 104 of which were previously classified as galaxies. 
The final selection of the 132 sources that were actually observed was made based on the availability 
of telescope time, in which we aimed to observed the largest galaxies first.
Note that for a sample chosen at high latitudes with 
matching 2MASS criteria, 96\% of the sources were previously catalogued in NED and all of these had measured 
redshifts.

The following 16 2MASS Working Database sources which we observed at \nan\ did not make it into the final 
2MASS source catalogue -- none of them were detected:   
002552+670925, 004129+600437, 022135+642636, 165630-324741, 
171245-341722, 173137-361819, 173535-141928, 174604-253240, 
180835-231312, 181804-044820, 181856-182017, 190102-140747, 
201300+354137, 2105131+493946, 2105142+493950, and 234810+610154.

\subsection{Arecibo sample} 
At Arecibo, we surveyed the ZoA galaxies ($|b| \leq 10\degr$) in the declination zone of 
+12$^{\circ}$ to +38$^{\circ}$. At the time of the Arecibo source selection (2000), the 2MASS Working 
Database did not yet cover the 0$^{\circ}$ to +12$^{\circ}$ declination range observable at Arecibo.
Due to constraints in the telescope time scheduling, the right ascension of the observed objects ranges from 
20$^{\rm h}$00$^{\rm m}$ to 06$^{\rm h}$30$^{\rm m}$, hence they mostly lie in the ZoA in the longitude range of 
about $\ell$ 170\dgr\ to 210\dgr\ plus a handful around 60\dgr.

In this zone, we observed 81 2MASS mostly low surface brightness sources, selected on a mean central $K_s$ 
surface brightness in the inner 5$''$ radius of $K_5 \geq $18 mag arcsec$^{-2}$, in the same manner as the 
2MASS LSB objects selected outside the ZoA by Monnier Ragaigne et al. (2003a).

All but one of 2MASS Working Database sources observed at Arecibo made it into final version of the 2MASS 
Extended Source Catalogue; the one source that did not (0639299+170558) was not detected.

It is worthwhile pointing out that most of the selected galaxies are on average much fainter and smaller 
than the \nan\ ZoA sample (see Fig.~\ref{RKAB} in Section~5) despite these galaxies being mostly in an area 
of low extinction and star density (roughly the Galactic anti-centre). The reason for this is that the majority 
of the brighter galaxies had been identified before by Pantoja et al. (1994) by visual examination of 
POSS E prints in their efforts to search for optical galaxy candidates  to trace the possible continuity 
across the Galactic plane of the southwestern spur of the Perseus-Pisces complex. They had then used Arecibo  
for follow-up \HI\ observations of the larger galaxies (Pantoja et al. 1997).

\section{Observations and data reduction } 
All radial velocities in this paper, both \HI\ and optical, are heliocentric and calculated according to the 
conventional optical definition ($V=c$($\lambda$-$\lambda_0$)/$\lambda_0$).

\subsection{\nan\ observations and data reduction} 
The \nan\ decimetric radio telescope, a meridian transit-type instrument of the Kraus/Ohio State design, 
consists of a fixed spherical mirror (300~m long and 35~m high), a tiltable flat mirror (200$\times$40~m), 
and a focal carriage moving along a curved rail track. Sources on the celestial equator can be tracked for 
about 60 minutes. The telescope's collecting area is about 7000~m$^{2}$ (equivalent to a 94-m diameter 
parabolic dish). Due to the E-W elongated shape of the mirrors, some of the instrument's characteristics 
depend on the declination at which one observes. At 21-cm wavelength the telescope's half-power beam width 
(HPBW) is \am{3}{5}~in right ascension, independent of declination, while in the North-South direction it 
is 23$'$ for declinations up to $\sim$20$^{\circ}$, rising to 25$'$ at $\delta$=40$^{\circ}$ and to
33$'$ at $\delta$=73$^{\circ}$, the northern limit of the survey (see also Matthews \& van Driel 2000). 
The instrument's effective collecting area and, consequently, its gain, follow the same geometric effect, 
decreasing correspondingly with declination. All observations for our project were made after a major 
renovation of the instrument's focal system (e.g., van Driel et al. 1997), which resulted in a typical 
system temperature of 35~K.

The initial observations were made in the period June - December 2002, using a total of about 350 hours 
of telescope time. A number of follow-up observations were made to check tentative detections in
late 2007 and in late 2008. We obtained our observations in total power (position-switching) mode using 
consecutive pairs of 40 seconds ON and 40 seconds OFF-source integrations. OFF-source integrations were 
taken at a position about 20$'$~E of the target position. Different autocorrelator modes were used for 
the observation of sources with previously known radial velocities and for velocity searches of objects 
of unknown redshift.

The autocorrelator was divided into one pair of cross-polarized receiver banks, each with 4096 channels 
and a 50~MHz bandpass, resulting in a channel spacing of 2.6~\kms. The centre frequencies of the 2 banks 
were usually tuned to 5600 \kms, for a velocity search in the $\sim$325 to 11,825 \kms\ range; for 8 
undetected galaxies, which are flagged in Table 2, the centre frequency was tuned to 5000 \kms, resulting 
in a $\sim$--275 to 11,225 \kms\ search range.

Flux calibration, i.e., the declination-dependent conversion of observed system temperatures to 
flux densities in mJy, is 
determined for the \nan\ telescope through regular measurements of a cold load calibrator and periodic 
monitoring of strong continuum sources by the \nan\ staff. Standard calibration procedures include 
correction for the above-mentioned declination-dependent gain variations of the telescope (e.g., 
Fouqu\'e et al. 1990). We also observed a number of calibrator galaxies throughout our observing runs, 
whose integrated line fluxes are on average 0.95$\pm$0.25 times our values (Monnier Ragaigne et al. (2003b).

The first steps in the data reduction were made using software developed by the \nan\ staff (NAPS, 
SIR program packages). With this software we averaged the two receiver polarizations and converted the 
flux densities to mJy. Further data analysis was performed using Supermongo routines developed by one of us 
(SES). With these we subtracted baselines (generally third order polynomials were fitted), excluding those 
velocity ranges with \HI\ line emission or radio frequency interference (RFI). Once the baselines were 
subtracted, the radial velocities were corrected to the heliocentric system. The central line velocity, 
line widths at, respectively, the 50\% and 20\% level of peak maximum (Lewis, 1983), the integrated flux 
of the \HI\ profiles, as well as the rms noise of the spectra were determined. All data were boxcar smoothed 
to a velocity resolution of 15.7 \kms\ for further analysis.

In order to reduce the effect of radio frequency interference (RFI) in
our observations, we used an off-line RFI mitigation program, which is
part of the standard \nan\ NAPS software package, see Sect. 3.3.

\subsection{Arecibo observations and data reduction} 
We observed a sample (see sect. 2.2) of 81 LSB galaxies in the ZoA from a pre-release version of the 
2MASS Working Database using the refurbished 
305-m Arecibo Gregorian radio telescope in November 2000 and January 2001, for a total of 
about 30 hours observing time. Data were taken with the L-Band Narrow receiver using nine-level 
sampling with two of the 2048 lag subcorrelators set to each polarization channel. All observations 
were taken using the position-switching technique, with the blank sky (or OFF) observation taken for 
the same length of time, and over the same portion of the Arecibo dish (as defined by the azimuth 
and zenith angles) as was used for the on-source (ON) observation. Each ON+OFF pair was followed 
by a 10s ON+OFF observation of a well calibrated, uncorrelated noise diode. The observing strategy 
used was as follows: First, a minimum of one 3 minute ON/OFF pair was taken of each galaxy, followed 
by a 10s ON/OFF calibration pair. If a galaxy was not detected, one or more additional 3 minute 
ON/OFF pairs were taken of the object, if it was deemed of sufficient interest (e.g., large diameter, 
known optical velocity).

The 4 subcorrelators were set to 25MHz bandpasses, and both
subcorrelators with the same polarization were set to overlap by 
5MHz. This allowed a wide velocity search while ensuring that the
overlapping region of the two boards was adequately covered. Two
different velocity searches were made -- first in the velocity range
--500 to 11,000 \kms\ and subsequently in the range 9,500 to 21,000
\kms\ (assuming the galaxy was not detected in the lower velocity range and observing time permitting). 
The instrument's HPBW at 21 cm is \am{3}{6}$\times$\am{3}{6} and the pointing accuracy is about 15$''$.

Using standard data reduction software available at Arecibo, the two
polarizations were averaged, and corrections were applied for the
variations in the gain and system temperature of the telescope with
zenith angle and azimuth using the most recent calibration data available at the telescope. Further data 
analysis was performed as mentioned above for the \nan\ data. A baseline of order zero 
was fitted to the data, excluding those velocity ranges with \HI\ line emission or radio frequency 
interference. Once the baselines were subtracted, the velocities were corrected to the heliocentric 
system, and the central line velocity, line widths at, respectively, the 
50\% and 20\% level of peak maximum (Lewis 1983), the integrated flux, as well as the rms noise of the 
spectra were determined. All data were boxcar smoothed to a velocity resolution of 14.3 \kms\ for analysis.

The stability of the chain of reception of the Arecibo telescope is shown by the observations we 
made of strong continuum sources and of a calibration galaxy with a strong line signal: the latter 
showed a $\pm$6\% standard deviation in its integrated line flux.

\subsection{Radio Frequency Interference (RFI)} 
As a consequence of their high sensitivity, radio astronomy telescopes are vulnerable to radio frequency 
interference (RFI), with signal strengths usually greatly exceeding those of the weak observed celestial 
radio sources. Broad-band RFI raises the noise level of the observations, while narrow-band RFI may mimic 
spectral lines like the \HI\ lines from galaxies that are being searched for in the present study. 
Besides external RFI, interference signals generated within the radio observatory, including the telescope 
system itself, may degrade the quality of the observations.

At \nan, where the renovated telescope had only recently been put back into operation at the time of 
the observations, persistent internal RFI occurred in the 3600--3800 and 4600--4900 \kms\ range and 
external RFI often occurred around 8300, 9000 and 10,500 \kms. The external RFI can be highly variable 
in time, and some occur in one polarisation only. 

At Arecibo an internal RFI source that wandered in frequency throughout the observed band, occurred 
regularly throughout the observing runs, besides the usual external RFI around 8300 and 15,000 \kms.

RFI signals with strengths that make the detection of faint \HI\ line
signals impossible in certain radial velocity ranges were present
during a significant fraction of the observations, both at \nan\ and
at Arecibo. At Arecibo, besides a hardware radar-blanker, no software
was available to identify and mitigate RFI signals. At \nan, we used
an off-line RFI mitigation program, as described in Monnier Ragaigne
et al. (2003b).

\section{Results} 
\nan\ observations of the following 12 2MASX sources could not be used due to extremely high rms 
noise levels ($>$15 mJy), most likely caused by strong nearby continuum sources: 04124692+3835153, 
04350092+5939419, 07392356-3221214, J08204513-3616164, 16171926-3740403, 16434955-3705384,
17504702-3116296, 18223005-0232233, 18340392-2524398, 20423202+4256315, 23044546+6004370, and 
23045989+6014030. We completely exclude these objects in the following tables, plots an discussions. 
Apart from these sources, the \nan\ rms typically ranges from about 2 -- 4 mJy for the detections 
with a few outliers,  whereas the non-detections spread more widely between 2 and 8 mJy
(see Tables 1 and 3). For the Arecibo observations, detections and non-detections all have 
quite low rms, hovering narrowly around 1 mJy  (see Tables 4 and 6). 

\begin{figure} 
\centering
\includegraphics[width=9.2cm] {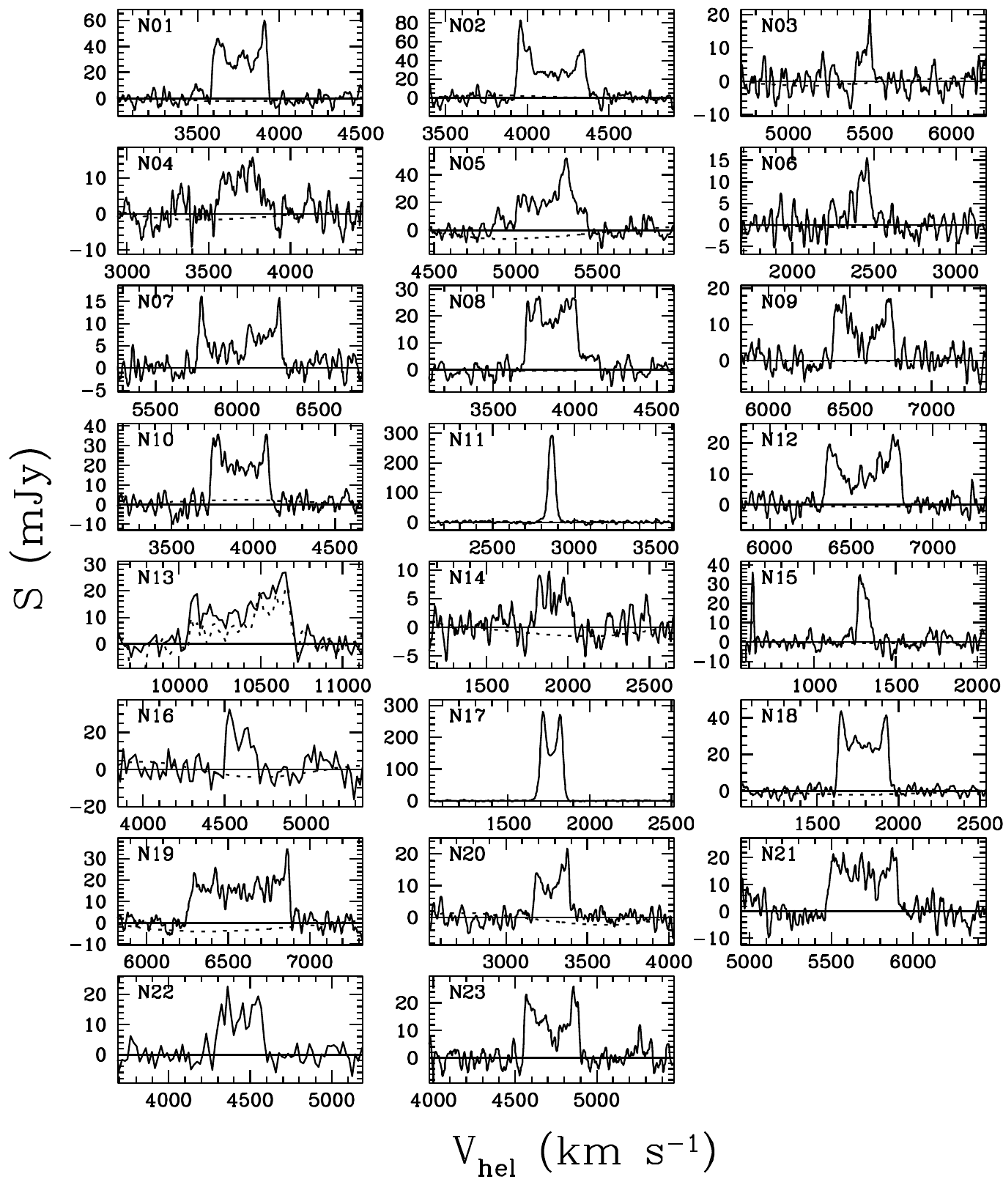}
\caption [] {\nan\ 21-cm \HI\ line spectra of the detected galaxies (see Table 1). Velocity 
resolution is 15.7 \kms. For object N13 two spectra are shown - the solid line represents the data taken 
in the direction of the target source, whereas the dashed line shows the data taken towards 
the galaxy 10$'$ South of it (see Sect. 4.3). For object N15, the spurious detection is actually of 
the galaxy ESO 371-27 which lies in the beam centred on the undetected target, ESO 371-26 
(see Sect. 4.1).}
\label{nancayspec}%
\end{figure}

Furthermore, we have not included the results for the 16 2MASS sources from versions of the 
Working Database sources that did not make it into the final 2MASS source catalogue (see Sect. 2.1), 
none of which were detected in \HI.

The resulting data are presented in the following figures and two sets of 3 tables for the \nan\ and 
Arecibo data. The \HI\ spectra of the objects detected at \nan\ and at Arecibo are shown in 
Figs.~2 and 3, respectively and the composite 2MASS $J H K_s$ images of all detected sources are 
displayed in Fig.~4. The \HI\ parameters of the  detected galaxies obtained with the \nan\ 
and Arecibo instruments are given in Table~1 and Table~4, respectively, including the main 2MASS 
parameters, and where available also the optical magnitudes and diameters. The latter were taken 
from the HyperLeda database (Paturel et al. 2003a), or retrieved from NED (then noted in parentheses). 
This is followed by Tables~2 and 5 which give the derived global properties of the galaxies. 
The non-detections are listed in Tables 3 and 6 with their main 2MASS parameters, Galactic extinction 
$A_B$, and \HI\ rms noise levels.

\begin{figure} 
\centering
\includegraphics[width=9.2cm] {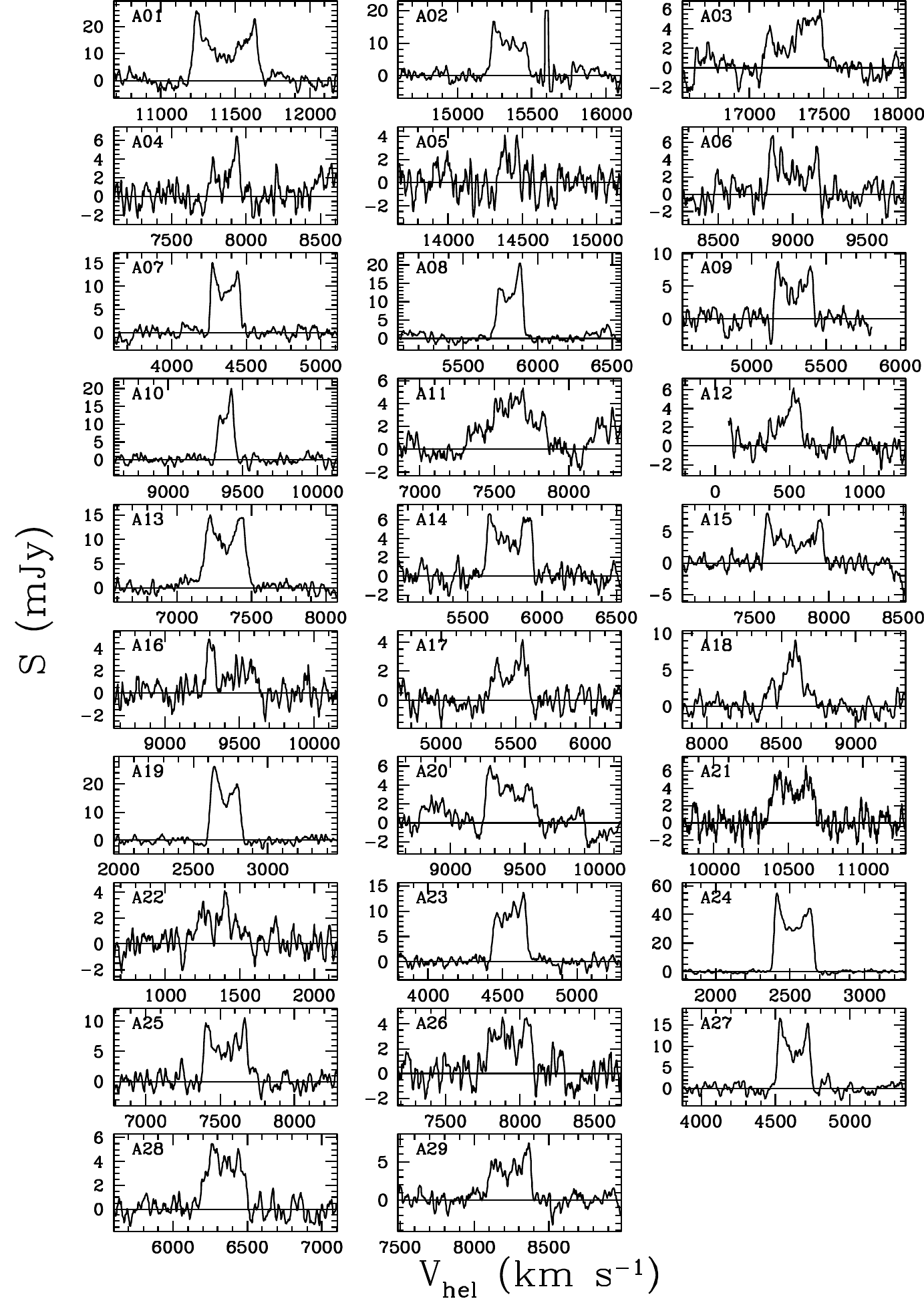}
\caption {Arecibo 21-cm \HI\ line spectra of detected galaxies (see Table 4). Velocity resolution 
is 14.3 \kms. }
\label{arecibospec}%
\end{figure}

The 2MASS data for these sources have all been updated to the final
release values. In a number of cases the Working Database values differed significantly, particularly 
because ellipse fits to galaxies can be quite unstable in the presence of multiple confusing foreground stars. 

The global \HI\ line parameters listed in the tables are directly measured values; no corrections 
have been applied to them for, e.g., instrumental resolution. Uncertainties $\sigma_{V_{HI}}$ in \VHI\ and
$\sigma_{I_{HI}}$ in \IHI\ can be determined following Schneider et al. (1986, 1990), as, respectively
\begin{equation} \sigma_{V_{HI}}=1.5(W_{20}-W_{50})X^{-1} (\kms)
\end{equation} 
\begin{equation} \sigma_{I_{HI}}=2(1.2W_{20}R)^{0.5}\sigma (\kms)
\end{equation} where $R$ is the instrumental resolution in \kms\ (see section 3), $X$ is the 
signal-to-noise ratio of a spectrum, which we define as the ratio of the peak flux density 
$S_{max}$ and the rms dispersion in the baseline, $\sigma$ (both in Jy). Following Schneider et al., 
the uncertainty in the $W_{20}$  and $W_{50}$ line widths is expected to be 2 and 3.1 times 
$\sigma_{V_{HI}}$.

\smallskip 

\noindent {\sl Description of all parameters listed in the Tables, in alphabetical order}:\\
\noindent
(1) {\it 2MASX J} is the entry number of a source in the final 2MASS
Extended Source Catalog, corresponding to the right ascension and declination
of the source centre in (J2000.0) coordinates.  \\
(2) $A_B$ is the Galactic $B$-band extinction in this direction in the Milky Way as estimated by Schlegel et al. (1998) \\
(3) $b/a$ is the infrared axis ratio determined from an ellipse fit to
the co-addition of the $J-$, $H-$, and $K_s$-band images\\
(4) $B_{T_{c}}$ is the total apparent $B-$band magnitude reduced to the RC3 system 
(de Vaucouleurs et al. 1991) and corrected for Galactic extinction, inclination and redshift effects 
(see Paturel et al. 1997, and references therein) \\
(5) $D=V_0/H_0$ is the galaxy's distance (in Mpc), where $V_0$ is its radial velocity (in \kms) 
corrected to the Galactic Standard of Rest and for infall towards various galaxy clusters in the 
local Universe, following Tonry et al. (2000), and a Hubble constant $H_0=75$ \kmsMpc \\
(6) $D_{25}$ is the diameter (in arcmin) at a visual surface brightness of approximately 25 \masq \\
(7) $I_{HI}$ is the integrated line flux (in \Jykms) \\
(8) $H-K$ and $J-K$ are the infrared colours within the $r_{K_{20}}$ isophotal aperture. Note that 
these colours are not corrected for extinction, hence reddened depending on the foreground dust 
column density by the amounts of $f_{H-K}=0.05A_B$ for $H-K$ and $f_{J-K}=0.12A_B$ for $J-K$ 
respectively \\
(9) $K_{20}$ is the total $K_s$-band magnitude measured within the $r_{K_{20}}$ isophotal aperture \\
(10) $k_{J-K}$ is the k-correction to the $J-K$ colour \\
(11) $l$ and $b$ are, respectively, the Galactic longitude and latitude of the source centre 
(in degrees) \\
(12) \LB\ is the $B$-band luminosity corrected for Galactic and internal extinction in solar units, 
for an assumed solar absolute magnitude of $5\fm48$ (Allen 1973) \\
(13) $L_K$ is the $K_s$ band luminosity of the galaxy in solar
luminosities within the $r_{K_{20}}$ isophotal aperture, for an assumed 
solar absolute magnitude of $3\fm31$ (Colina \& Bohlin 1997) \\
(14) $\frac{{M}_{baryon}}{M_{dyn}}$ is the ratio of the combined \HI\
and stellar baryonic mass as a fraction of the total dynamical mass, where
$M_{baryon}$=0.8\LK+1.4\MHI\ (McGaugh et al. 2000 or 2003) \\
(15) $M_{dyn}$ is the dynamical mass (in \Msun) estimated from the rotation speed and 
the $K_s$-band radius, $M_{dyn}=v_{rot}^2 r_{K_{20}}/G$ \\
(16) \MHI\ is the total \HI\ mass (in \Msun), \MHI=2.356 $10^5 D^2$ \IHI\\
(17) \MHILK\ is the ratio of the total \HI\ mass to the $K_s$-band luminosity in solar units \\
(18) {\it No} gives the source number used in Figs. 2--4 \\
(19) {\it Other Name} is the entry number in another major galaxy catalogue \\
(20) {\it PGC No} is the entry number in the Principal Galaxy Catalogue (Paturel et al. 1989) \\
(21) $r_{K_{20}}$ is the radius (in arcsec) at a surface
brightness of 20 \masq\ in the $K_s$ band (in arcsec in Tables 1 and 4, in kpc in Tables 2 and 5) \\
(22) $rms$ is the rms noise level or $\sigma$ in the \HI\ spectrum (in mJy) -– if two numbers are 
given, the first is for the low-velocity search and the second for the high-velocity one (see Sect. 2.2) \\
(23) $S_{max}$ is the peak flux density of the line (in mJy) \\
(24) $V_{50}$ is the heliocentric central radial velocity of a line profile (in \kms), in the 
optical convention, taken as the average of the high and low velocity edges of the \HI\ profile,
measured at 50\% of peak flux density \\
(25) $v_{rot}$ is the rotation speed corrected for inclination $i$; $v_{rot}=W_{50}/2sin(i)$ for
sin($i$)$<$0.2, for galaxies with higher inclinations we assumed $v_{rot}=W_{50}/2$  \\
(26) $W_{50}$ and $W_{20}$ are the profile's velocity widths 
(in \kms) at 50\% and 20\% of peak maximum, respectively.

\subsection{Comparison with published \HIit\ data} 
In the literature we found the following 10 \HI\ detections of sources we observed (see also 
section 4.1 and Table 7): 4 made at Arecibo (A01, A19, A21 and A23), 1 at Effelsberg (N12), 
1 at \nan\ (N10), and 4 at Parkes (A24, N11, N17 and N18). Excluding the Effelsberg 
data for N12, which are affected by RFI, and the Parkes data for A24, which appears to have been 
resolved by the Arecibo beam, we find a good overall agreement between the global profile parameters 
measured by us and taken from the literature: the mean of the absolute values of the differences is 
7$\pm$8 \kms\ in $V_{HI}$ and 7$\pm$6 \kms\ in $W_{50}$, and our $I_{HI}$ fluxes are on average 
1.0$\pm$0.2 times the literature values.

\subsection{Notes on individual galaxies} 
In order to identify galaxies within the telescope beams that might give rise to confusion with the 
\HI\ profile of the target galaxy, we inspected 2MASS and DSS images centred on the position of each 
clearly or marginally detected source, over an area of 12$'$$\times$36$'$ and $8'$$\times$8$'$ 
($\alpha$$\times$$\delta$) for the \nan\ and Arecibo data, respectively, and queried the NED and HyperLeda 
databases for information on objects in these areas -- the results for objects whose data are likely 
to be confused are put in brackets in the Tables. The data listed below were preferentially taken from 
the mean values listed in HyperLeda, unless otherwise indicated.

\begin{figure} 
\centering
\includegraphics[width=9cm] {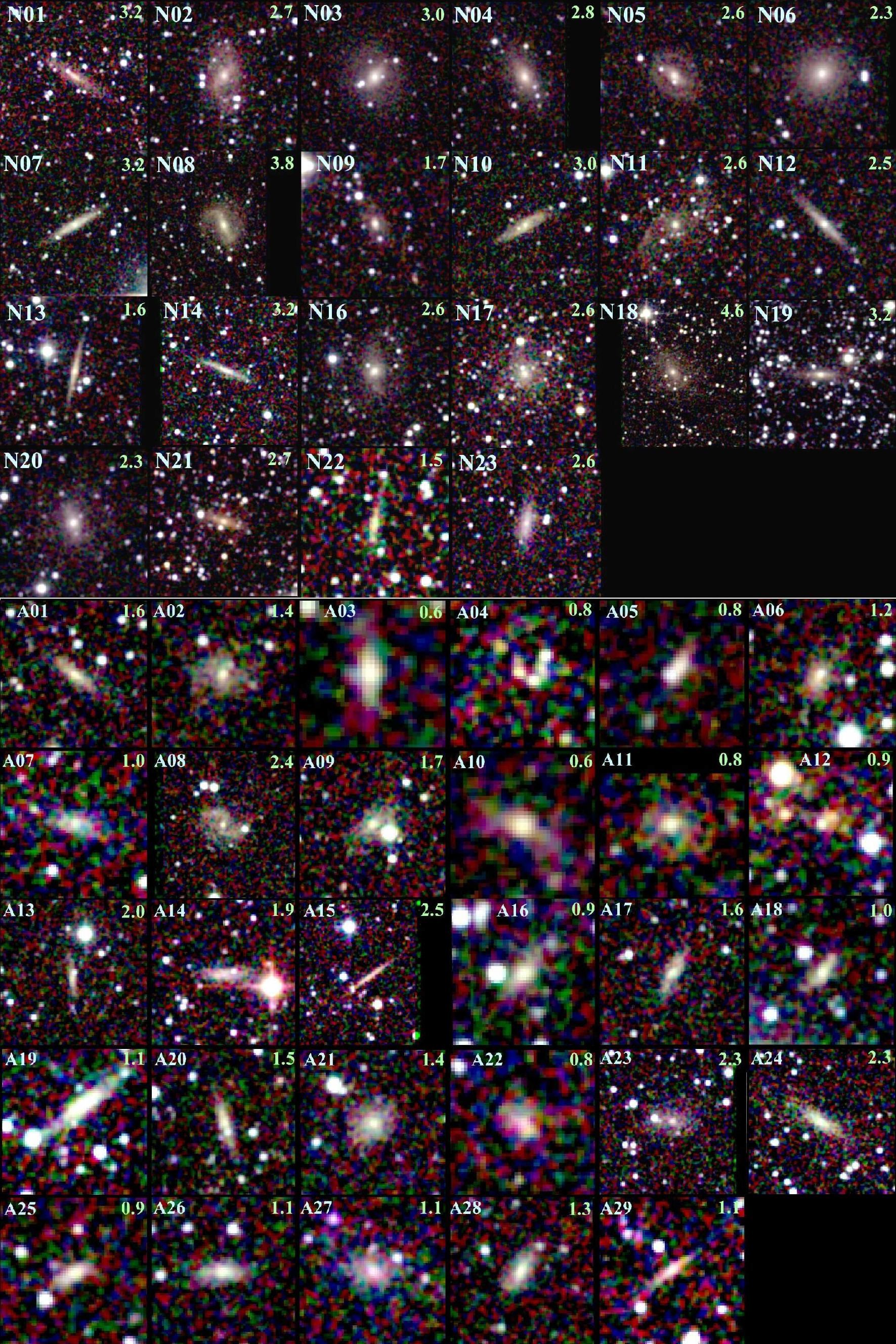}
\caption {Composite 2MASS composite $J H K_{s}$-band colour images of the ZoA galaxies detected in 
the \HI\ line at \nan\ (top panel) and at Arecibo (lower panel). The galaxy identifications (see 
Tables 1 and 4) are indicated in the top-left corner of each image and the image size in arcmin 
in the top-right corner. }
\label{zoadets}%
\end{figure}

{\sl 2MASX J04212943+3656572} (=N06): it appears also to have been detected at \nan\ by Theureau et al. 
(1998) while pointing towards the nearby galaxy NGC 3016, which lies on the \nan\ HPBW edge. 
NGC 3016 has a much higher redshift of 5900 \kms\ (Pantoja et al. 1997; Huchra et al. 1983). 
The galaxy NGC 3019, which also lies on the \nan\ HPBW edge, has a similarly high redshift of 
5664 \kms\ (Fisher et al. 1995; Pantoja et al. 1997; Takata et al. 1994; Theureau et al. 1998).

{\sl 2MASX J04215207+3607373} (=UGC 3021): classified as an elliptical galaxy, and therefore not 
expected to be gas-rich. Its optical velocity is 6238$\pm$60 \kms\ (Huchra et al. 1983). 
It was not detected by us at \nan\ with an rms of 2.8 mJy, nor at Arecibo by Pantoja et al. 
(1997), whose mean rms is 1.3 mJy. It has a possible companion superimposed on it, hence the 
two PGC entries.

{\sl 2MASX J04514426+3856227} (=N10): its optical redshift of 3913$\pm$60 \kms\ (Saunders et al. 2000) 
is in agreement with our \nan\ detection and that of Paturel et al. (2003b).

{\sl 2MASX J05315137+1517480} (=A08): an optical velocity of 5000$\pm$60 \kms\ (Tully 2002, 
private communication) was listed in Hyperleda after our Arecibo detection was made at 5812 \kms.

{\sl 2MASX J06301575+1646422} (=A24): detected in \HI\ at Parkes as HIZOA J0630+16 (Donley et al. 2005) 
with a three times higher line flux than our Arecibo detection. This indicates that the source has a
quite extended LSB disc, which has to be much larger than its $\sim$1$'$ extent on the composite 2MASS 
$J H K_{s}$-band image, given the respective Arecibo and Parkes HPBW of \am{3}{6} and \am{14}{4}.

{\sl 2MASX J07300453-1833166}: this source appears to be part of a
Galactic \HII\ region, together with other nearby 2MASS sources and object ESO559-N015.
 
{\sl 2MASX J08080461-1452387} (=N12): detected in \HI\ at Effelsberg
by Huchtmeier et al. (2005) at 6679 \kms, which is significantly higher than our value of 6575 \kms. 
The published spectrum shows a narrow peak at 7050 \kms, however, which does not occur in our spectrum 
and thus appears due to RFI. We re-estimate the Effelsberg profile parameters as listed in Table 7, 
which are in agreement with ours. 

{\sl 08544150-3248590} (near N15; =ESO 371-27): after our \nan\ detection was made at 1302 \kms\ 
(consistent with the \nan\ spectrum of Chamaraux et al. 1999, taking into account beam attenuation) 
an optical redshift of 2198 \kms\ was published (Wegner et al. 2003), which shows that we actually 
detected another galaxy within the beam, ESO 371-27, which was also detected at Parkes (Doyle et al. 2005). 
All \HI\ redshifts of ESO 371-27 are consistent with the optical value of 1313 \kms\ 
(Karachentseva \& Karachentsev 2000). As the Parkes line flux is significantly higher we have used 
it for calculating the total \HI\ mass. The detected galaxy is very LSB and it does not have an entry 
in the 2MASS catalogue. 

{\sl 2MASX J18153013-0253481} (=N18): Inspecting the on-line Parkes \HI\ spectrum of HIPASS J1815-02 
(Meyer et al. 2004) we found that the HIPASS redshift of 1664 \kms\ listed in NED is actually that of 
one of the profile's two peaks; the correct value is 1788 \kms.

{\sl 2MASX J21135161+4255323} (=N22): Although the targeted source (21135100+4257568) did not make 
it into the final 2MASS catalogue we made a clear detection towards this position at \nan. We have 
assumed this to be of 2MASX J21135161+4255323, located \am{1}{3} towards the south.

\subsection{Unusual Galaxies} 
{\sl 2MASX J05422061+2448359} (=A12): Curiously, the galaxy that is nearest ($v=521$\kms) and lowest 
in \HI\ mass (log\MHI= 7.16), is also the reddest in our sample. Even after correcting for the 
reddening of the source ($A_B=4.42$), the extinction-corrected $(J-K)^o$ colour ($1\fm32$) is more 
likely that of an extremely 
old elliptical than a small \HI-rich dwarf. It is possible that the extinction in this direction is 
substantially larger than estimated by Schlegel et al. (1998). If the reddening were large enough, 
however, to give this object a more-typical $(J-K)^o$ colour, the extinction correction would imply an 
extremely large stellar mass.

Another obvious possibility is confusion with a red star. Apart from the entry in 2MASX, the 2MASS 
point source catalog lists two very nearby stars at 1\arcsec and 5\arcsec distance respectively, 
where the nearer stellar counterpart might be one and the same object. The source centred on the 
targeted 2MASS position  actually looks more like a point source with more typical colours of a star 
(strong in $H$), whereas the slightly more offset 2MASS source looks more like a fuzzy reddened 
galaxy candidate (see Fig.~\ref{zoadets}; slightly to the NW of the central source: 2MASS 05422020+24483880). 
In either case the colour of the extended object will be substantially contaminated by the nearby star.

A third possibility is that neither of these objects is the counterpart of what must be a highly 
obscured late-type spiral galaxy (narrow Gaussian profile, low \HI\ mass), but that the \HI\ 
detection originates from the nearby completely obscured infrared source IRAS~05393+2447 -- 
or even some other invisible galaxy.

{\sl 2MASX J08170147-3410277} (=N13): This thin edge-on spiral galaxy has a radial velocity (10,369 \kms), 
\HI\ mass ($4.6 \cdot 10^{10}$ \Msun), and inclination-corrected rotation velocity (314 \kms) similar to 
that of the very \HI-massive disc galaxy HIZOAJ0836-43, discovered by Donley et al (2006; $V$=10,689 \kms, 
\MHI$=7.5 \cdot 10^{10}$ \Msun, and $v_{rot}$=305\kms). The latter has about twice the  estimated total 
dynamical mass of  N13 (1.4 vs. $0.6 \cdot 10^{12}$ \Msun), comparable to that of the most massive known 
disc galaxies such as giant LSB galaxy Malin 1. It is, however, a NIR luminous star-bursting galaxy 
(Cluver et al. 2008) with quite distinct properties from giant LSBs.

Because such high \HI\ mass galaxies are (a) extremely rare (they are only being formed now and 
their properties poorly known) {\sl and} (b) our clear \nan\ detection with a 9$\sigma$ peak 
flux density of 27 mJy could not be found back in the deep Parkes ZoA survey data cube ZOA252  
(http://www.atnf.csiro.au/research/multibeam/release/), we looked at this 
galaxy and detection in further detail. We first looked for possible companions which might have 
contributed to the broad signal by inspecting all bands of the digitized sky survey within the 
\nan\ beam as the extinction is relatively low ($A_B \sim 2$mag). We found a previously uncatalogued 
galaxy of similar surface brightness about 10$'$ south of the target, at $\alpha=
08^{\rm h}$16$^{\rm m}$\tis{58}{6}, $\delta= -34^{\circ}$20$'$\as{0}{24}. It appears on the $B, R$ 
as well as the $IR$ images as a smaller (roughly half an arc minute in diameter) face-on spiral 
with a smallish bulge.

With an angular N-S separation of 10$'$ corresponding to 0.45 times the instrument's HPBW, observations 
towards both objects allow us in principle to disentangle their \HI\ profiles, if their angular diameters 
are sufficiently small. We therefore (re)-observed both the sources N13 North and N13 South, to a 
similar low noise levels (see Fig. 2). The velocity range and central velocity of both profiles are the same, 
whereas the profile towards the southern object has 76\% of the \HI\ mass measured towards the northern 
one and its \HI\ is mainly concentrated in the high-velocity peak of the double horned profile. 
This appears to indicate the presence of a single, relatively large source towards N13 North whose 
receding half is towards the south.

We attempted to reconstruct the spectra of the hypothetical \HI\ sources N13 North and South, assuming 
that their sizes are significantly smaller than the telescope beam. As they are separated by about half 
a HPBW, this would imply that an observed spectrum is due to the target plus half of the emission of 
the other galaxy. This exercise showed that N13 South is not a significant source of \HI\ in itself.

Whereas our provisional conclusion is that N13 appears to be an extended, very massive \HI\ galaxy, it is 
clear that only ATCA imaging observations will allow a firm conclusion whether it has a very extended 
lopsided \HI\ disk and belongs to the class of extremely massive spiral galaxies.

\section{Discussion} 
\subsection{Detection rate} 
The detection rate is quite low. Excluding the 16 Working Database sources that did not make it into the 
final 2MASS Extended Source Catalog (Sect. 2.1), only 22 of the 116 observed sources (19\%) were detected 
at \nan, excluding the detections of spurious, untargeted galaxy in the beams of N15 and N22. This value is only 
a bit higher (21\%) if we exclude the 12 strongly continuum perturbed spectra (Sect. 4) -- generally a problem 
at very low latitudes. The on average 3.5 times higher sensitivity Arecibo observations resulted in a detection 
rate of 36\% (29/81), only 1.9 times higher (36/19) than that at \nan.
Criteria for improving the selection of likely 2MASX candidate ZoA galaxies are discussed in Sect. 5.2
and 6.

There are various reasons -- partly different for the \nan\ and Arecibo observations -- for the low 
detection rate. For both samples, no morphological type criterion was introduced when selecting the 
target objects, i.e. the sample includes both red gas-poor galaxies as well as  blue gas-rich ones. 
Being NIR selected, the bias against the more blueish gas-rich galaxies is quite strong, stronger 
than, for instance, for optical selected samples. This bias is even more extreme for the relatively 
shallow and low-resolution 2MASS survey (as discussed in Section~1), which is hardly sensitive to LSB 
galaxies. It is exacerbated for ZoA galaxy candidates because of the increasing loss of low-surface 
brightness features {\sl and} the selective reddening, which results in an even stronger bias towards 
redder, higher surface brightness early type galaxies or bulges of spirals. Given that optical 
spectroscopy of these optically heavily or completely obscured 2MASS galaxies is hardly possible, it 
should be noted that despite this relatively low detection rate, \HI\ observations of galaxy candidates 
still remain the most efficient tool in mapping large-scale structures across the ZoA.

\begin{figure} 
\centering
\includegraphics[width=9cm] {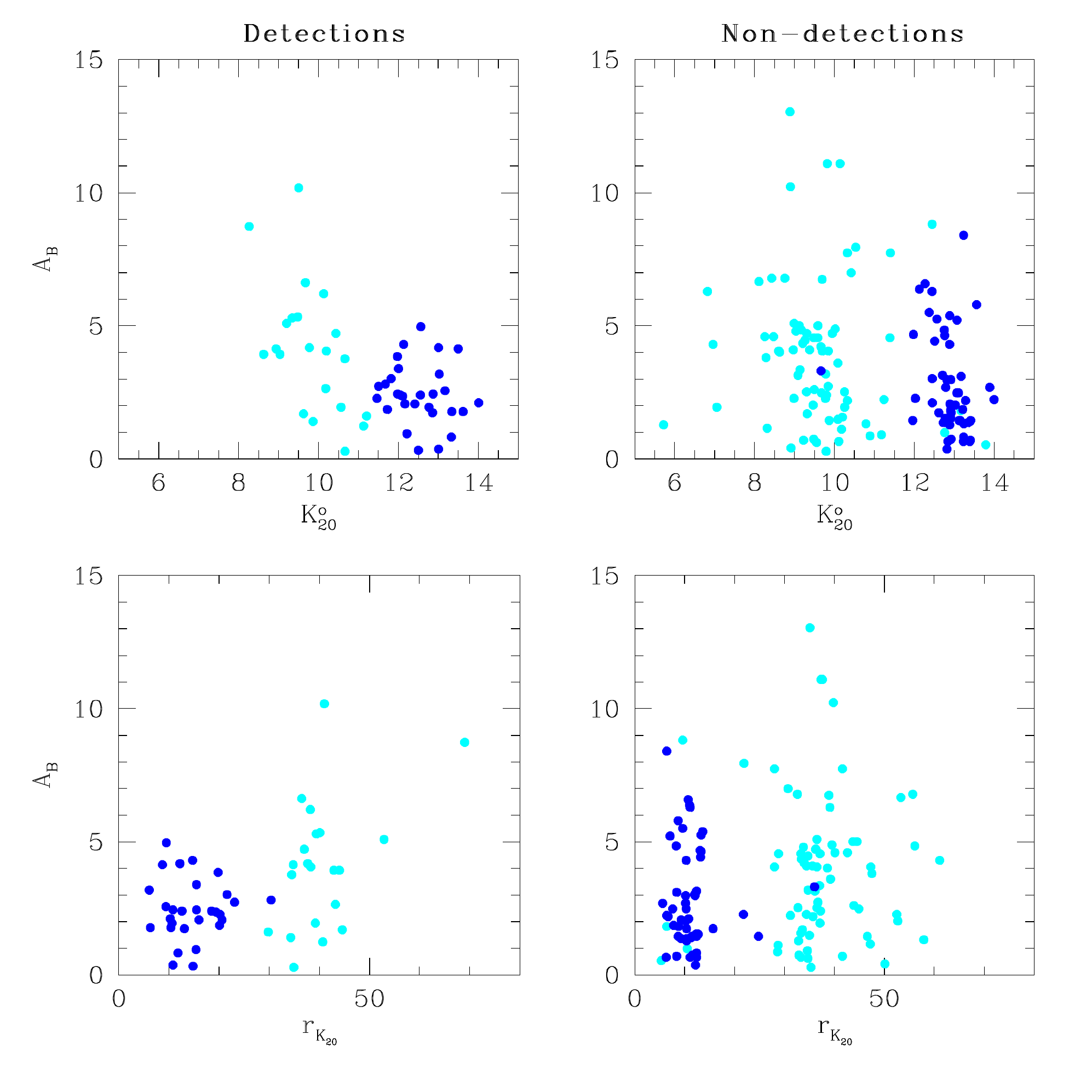}
\caption {Plots of the extinction-corrected total $K_s$-band magnitude, $K_{20}^0$, and the 
$r_{K_{20}}$ isophotal aperture as function of the galactic extinction $A_{B}$. The two left-hand 
panels show the \HI\ detections, the two right-hand ones the non-detections. Objects observed at 
\nan\ are shown in light blue and those observed at Arecibo in dark blue. \HI\ detections are shown 
in the two left-hand panels, non-detections on the right. The crosses mark typical 2MASS galaxy 
colours (Jarrett 2003}
\label{RKAB}%
\end{figure}

Some of the properties of the two galaxy samples (\nan\ and Arecibo) and the differences between them, 
as well as between detections and non-detections are apparent from Fig.~\ref{RKAB} which shows plots of 
the extinction-corrected $K^0_{20}$ magnitude versus optical extinction $A_B$ (top panel) and of radius 
(at $r_{K_{20}}$) versus $A_B$ (bottom panel) for both detections (left panels) and non-detections 
(right panels). The \nan\ sample is indicated in light blue and the Arecibo one in dark blue. 

It is obvious that the Arecibo galaxy sample contains a fainter subset of galaxies compared to the 
\nan\ sample, i.e. roughly ranging from $12 - 14$ mag compared to $6 - 11$ mag, as well as smaller 
galaxies ($\la 30\arcsec$ versus $\ga 30\arcsec$). This is due to the fact that many of the larger 
galaxies in the low extinction Arecibo area were already identified optically and observed with 
Arecibo by Pantoja et al. (1994, 1997). Although the \nan\ galaxies are considerably brighter (both 
observed and extinction-corrected), they are traced deeper into the Galactic dust layer. The lower 
detection rate of the brighter \nan\ galaxies is purely the result of the lower sensitivity.

Overall there seems no marked difference between the locus of points for detections and non-detections, 
substantiating once more that \HI\ observations are unaffected by dust-extinction. However, two trends 
seem to stand out: for extinction values $A_B$ above about 5 mag all 2MASX Arecibo sources are undetected. 
This implies that these apparently small, highly obscured non-detections must be very distant early-type 
galaxies. 

Secondly, it seems surprising that none of the brightest \nan\ objects (extinction-corrected) were detected. 
This must be due to confusion in 2MASX with Galactic sources or misclassification of  galaxies due to
blending of sources. To verify this claim, and learn from this pilot project, we investigated the brightest 
sources individually.

The two brightest sources ($K^o_{20}=1\fm82$, $5\fm72$; of which the brighter lies beyond the boundaries of 
Fig.~\ref{RKAB}) are globular clusters with the first one having colours inconsistent with a galaxy (see 
also Fig~\ref{colours} in the next section) and the second being borderline. But the globular cluster 
morphology is obvious enough in all optical and NIR wavebands.

The $3^{rd}$ galaxy in the list of decreasing extinction-corrected brightness $K^o_{20}$ (with 6\fm82) is 
also an unlikely galaxy. It has the extreme colour of $(H-K^o)=4\fm34, (J-K)^o=0\fm27$, and is clearly 
heavily contaminated, in  particularly in the $H$-band, by a bright nearby star. The remaining feature 
does not look like a galaxy candidate. 

The $4^{th}$ galaxy candidate (6\fm96) is a more difficult case. The extinction-corrected colours are 
compatible with this being a real galaxy (0\fm29, 0\fm91) and it certainly has the appearance of a 
galaxy, looking like a spiral with a clear bulge and LSB disc -- or a central bright star with some 
nebulosity around it. The optical does not provide clarification either. However, if it were a large 
spiral galaxy (which then should have $B^o\sim 9\fm0$) it should be visible on the blue (IIIaJ) and 
red sky survey plate at an extinction of 'only' $A_B=4\fm5$. There is no evidence for that at all 
on the respective optical sky survey plates. Hence we doubt it to be real. 

The $5^{th}$ object ($K^o_{20}=6\fm99$) is clearly disqualified based on extinction-corrected colours 
alone (0\fm23, --0\fm06). It is also improbable given the thick dust layer ($A_B$=24\fm69) 
through which it has been observed. However, it looks like a possible galaxy on the 2MASX image 
with a bright centre and LSB disc, despite its blue $(J-K)^o$ colour. Here, the optical image 
provides additional help. It shows the target to be a nebular object with a star at its centre. 
Interestingly this is also a strong radio continuum source.

The $6^{th}$ galaxy in the list (with 7\fm05) is a very obvious and bright galaxy at only intermediate 
extinction levels in the Puppis area. It is catalogued as ESO 430-G028 in the ESO Uppsala and ESO--LV 
catalogues (Lauberts, 1982; Lauberts \& Valentijn 1989). But being an S0 galaxy, it is clear why it was 
not detected with our \HI\ observation. In fact, it does not even have a published redshift yet.

The colours (--0\fm19; --0\fm49)and high extinction level extinction
($A_B=18\fm8$) of the $7^{th}$ galaxy ($K^o_{20}=7\fm18$) make it completely unrealistic for this 
to be a galaxy. Again visual inspection of both the NIR and optical image confirms this. The object 
looks like a star forming region with resolved individual stars and some reddish nebulosity around 
it in 2MASS whereas the optical shows none of these resolved stars, only a few stars (probably 
foreground) on a nebula (probably a reflection nebula). This also is a strong radio source.
 
The subsequent objects further down on this magnitude list ($K^o_{20} > 8\fm0$) are mostly definite 
galaxies, with only a minor fraction of uncertain or unlikely galaxies. The majority of these unlikely 
galaxies have extinction-corrected colours that are incompatible with these being galaxies obscured 
-- and reddened -- by the Milky Way, as they are quite blue.

So the relative lack of detections of intrinsically bright 2MASX objects in our sample can be understood.

\subsection{Near-infrared colours of the galaxy sample}  
\begin{figure} 
\centering
\includegraphics[width=9cm] {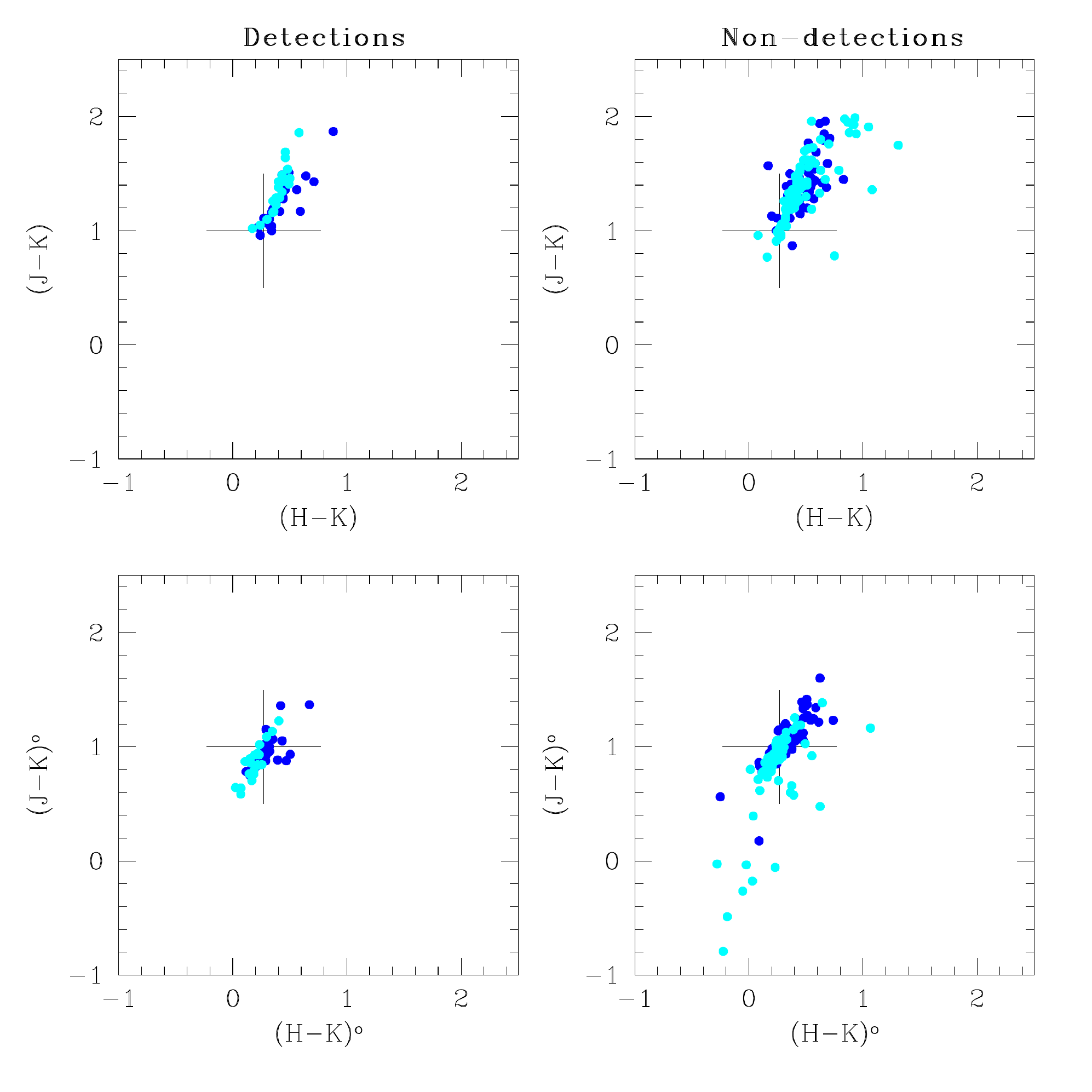}
\caption {Plots of the observed (top) and extinction-corrected (bottom) colours $(J-K)$ versus 
$(H-K)$, for detections (left panels) and non-detections (right panels). Objects observed at 
\nan\ are shown in light blue and those observed at Arecibo in dark blue. }
\label{colours}%
\end{figure}

Figure~\ref{colours} displays the observed and extinction-corrected $H-K$ vs. $J-K$ colour-colour 
diagram for the 2MASS galaxy sample. The cross gives an indication of the mean colours 
for unobscured galaxies as given in Jarrett et al. (2003), namely $H-K=0\fm27$ and $J-K=1\fm00$, 
which has a notoriously low dispersion about the mean. Only the late type Sd to Sm spirals have 
clearly lower colours (dipping to 0\fm7 and 0\fm15 for $J-K$ and $H-K$, respectively). Such 
bluish low surface brightness galaxies will, however, have entered our sample only in small numbers 
for the earlier mentioned reasons.

A comparison between the four panels allows some interesting observations. The colour plots uncorrected 
for extinction (top) are, as expected, quite similar as the extreme blueish objects were eliminated 
from the observing list, but no discrimination against reddened objects was made because of the 
foreground Galactic dust reddening. When correcting the colours for extinction, the detected galaxies 
(left bottom panel) fall quite nicely within the expected colour range for galaxies. The data points 
are, however, not distributed in a Gaussian cloud but rather in a more elongated linear distribution 
along the line of reddening. This can be explained by an over- or underestimate in the adopted extinction 
corrections. The likelihood for an overestimate is larger (independently confirmed in section 5.5.) 
given that a larger fraction of the galaxies lie above the mean of unobscured galaxies as found in 
Jarrett et al. (2003) in the extinction-corrected colour-colour diagram.

This also holds for a large fraction of the non-detections. However, there are over a dozen extremely 
blue objects (three blue objects fall beyond the axes limits displayed in Fig.~\ref{colours}, with 
the most extreme having colours of  $(H-K)^o=-0\fm64$ and $(J-K)^o=-1\fm68$, and one with extreme 
red $(H-K)^o$ colours). These objects clearly cannot be extragalactic.

The two most extreme sources (2MASX J07300453-1833166 and 2MASX J07300594-183254) lie at extinction 
levels of $A_B\sim 80$mag (according to the Schlegel et al. (1998) maps), which even in the $K_s$ 
band implies over 7 magnitudes of extinction. 
Indeed, visual inspection of the NIR images as well as the optical images, find these two objects 
to be stars within \HII\ regions, i.e. point-like objects with some fuzziness around them that emit 
strongly in the $H$-band. The same holds for most of the other objects. They are either stars that 
are located in or associated with an \HII\ region, or some filamentary Galactic nebulosity (e.g. 
2MASX J23352762+6452140) generally visible in both the optical and infrared, or the image is heavily 
contaminated by a very bright nearby star, and the resulting galaxy classification highly uncertain 
(e.g. 2MASX J20491597+5119089 with colours of $(H-K)^o=4\fm34$ and $(J-K)^o=0\fm27$).

Indeed, the galaxy with the bluest colours that was actually detected in \HI\ (N17; with 0\fm07 
and 0\fm59) is the first of the targeted objects in a list of increasing $(J-K)^0$ colour that 
actually has the appearance of a galaxy on  NIR and optical images, and its properties. This 
observation is independently confirmed by Jarrett who visually examined all 2MASX sources within 
$\pm10$\dgr\ of the Galactic Plane (priv. comm.) and classified most of these objects bluer than 
N17 as non-galaxies.

The lesson learned from these results indicate that the \HI\ detection rate of 2MASS-selected 
ZoA galaxies can be significantly improved if, in addition to the exclusion of galaxy candidates 
with observed blue NIR colours $J-K$ and $H-K$, 2MASX objects are also excluded that have 
{\sl extinction-corrected colours} $(J-K)^0 < 0\fm5$ and $(H-K)^0 < 0\fm0$.

The easiest way to apply such corrections is by systematically using the DIRBE extinction maps. 
Although we do find (see section 5.5) -- like many others (Schr\"oder et al. 2007 and references 
therein; Cluver et al. 2008) -- that when taking the DIRBE extinction measures at face value in 
southern ZOA studies, these seem overestimated by about 15\% to 50\% (Nagayama et al. 2004; 
Schr\"oder et al. 2007; Tagg 2008; Cluver 2008).

This overestimate will, however, have only a minimal effect on the extinction-corrected colours, 
as the selective reddening of $f_{J-K}=0.12A_B$ for $J-K$ and $f_{H-K}=0.05A_B$ for $H-K$ 
will reduce the colours by a relatively low amount. For instance, a reduction by an intermediate 
overestimate of 30\% of the DIRBE extinction values and an intermediate to high ZoA dust column 
density of $A_B=5\fm0$ results in a decrease of the extinction-corrected colour by the relative 
modest amount of $(J-K)^0=0\fm36$ and $(H-K)^0=0\fm07$. 

If such an extinction-corrected colour limitation is then followed by visual examination of the 
individual and combined $JHK_s$ images, partly in combination with optical images to help eliminate 
Galactic objects, cirrus, filaments and blended images, then a fairly high \HI\ detection rate should 
be guaranteed given that spiral galaxies are generally more common than early type galaxies -- even 
if we cannot discriminate against morphological type with NIR colours. 

\subsection{Global properties of the detected galaxies}  
\begin{figure} 
\centering
\includegraphics[width=9.2cm] {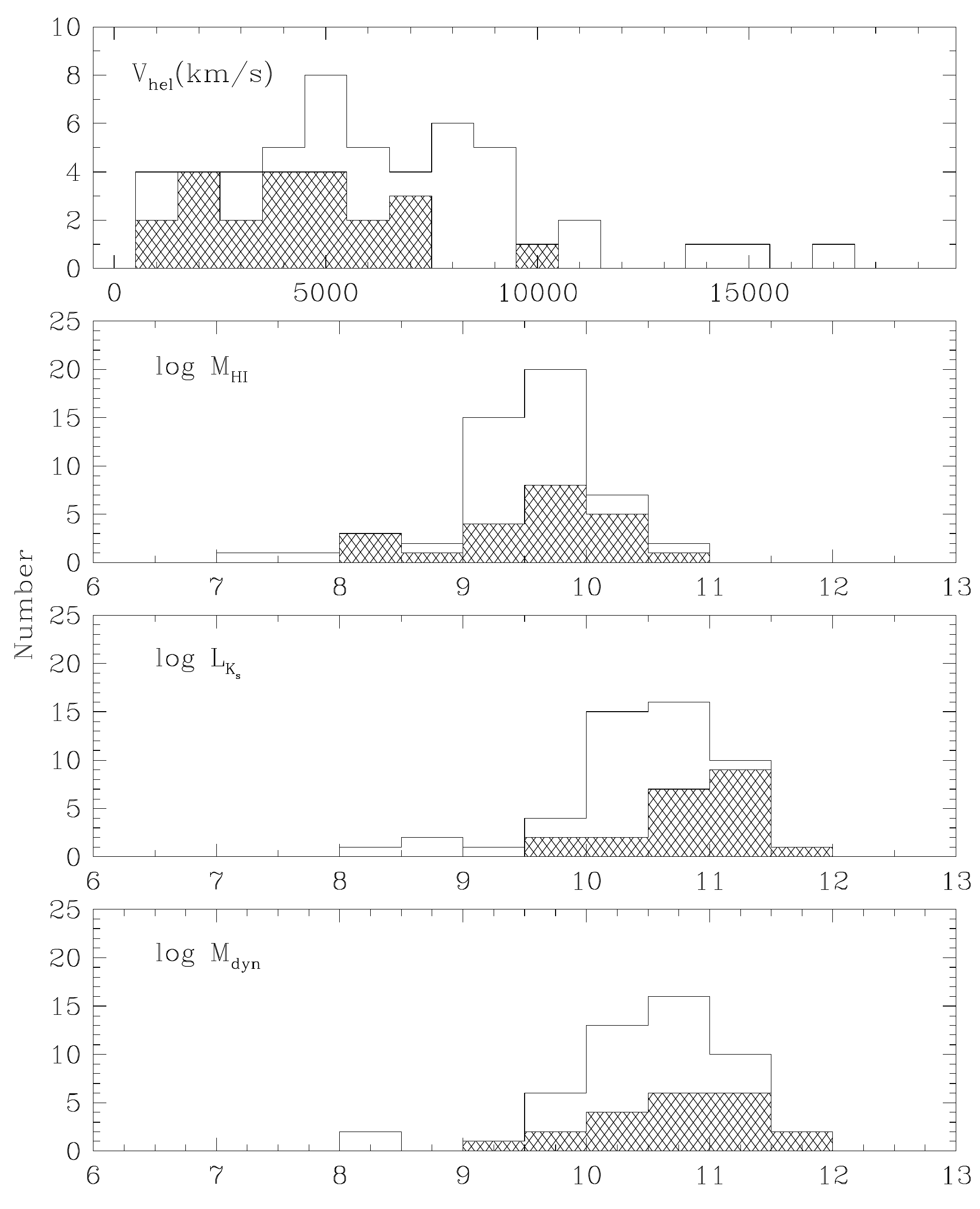}
\caption {Histograms of the distribution of the radial velocity, the total \HI\ mass $M_{HI}$ in 
\Msun, Galactic extinction-corrected $K^o_{20}$ band luminosity $L_{K}$, in \LsunK, and the dynamical 
mass $M_{\rm dyn}$, also in solar units, of the galaxies detected at \nan\ (shaded area) and Arecibo.}
\label{histo}%
\end{figure}

In the following, we have a brief look at the distribution of the global properties of the detected 
galaxies such as radial velocity, $K_s$-band luminosity $L_{K_{20}}$, total \HI\ mass $M_{HI}$, and 
dynamical mass $M_{dyn}$ -- see the resulting histograms  in  in Fig.~\ref{histo}. The hashed histograms 
denote the \nan\ observations and the clear ones the Arecibo detections.

%
The velocity distribution of the \nan\ detections shows galaxies out to about 7000\kms, but beyond that 
its efficiency drops quite rapidly toward the velocity search limit of $V_{hel}=11'825$\kms. This 
is very similar to the systematic southern Parkes HIPASS ZoA survey  (Kraan-Korteweg et al. 2005; see 
their Fig. 2), which has similar instantaneous velocity coverage with a slightly lower sensitivity limit 
(rms=6 mJy), except for the prominent peaks  in their survey due to the crossing of the Hydra/Antlia 
filament and Great Attractor Wall (centred at about 3000 and 5000\kms\ respectively). This suggests that 
a systematic \HI\ follow-up of 2MASX ZoA objects -- with the current setup of the \nan\ pilot project 
observations -- would be quite complementary to the southern ZoA efforts. Apart from the systematic 
(``blind'') ZoA ALFA survey undertaken for the declination range visible with the Arecibo telescope 
(e.g. Henning et al. 2008; Springob 2008), no such efforts are currently being undertaken for the northern ZoA. 

The present Arecibo detections have a higher mean velocity Peak -- the majority lie between 5000 – 10,000\kms\ 
with a handful of galaxies up to 18,000\kms. The fact that very few galaxies have been detected at 
low velocities is -- as mentioned in section 5.1 -- due to the work of Pantoja et al. (1997) who 
have already observed most of the (optically) larger galaxies at Arecibo in their efforts to map 
the southwestern spur of the Perseus-Pisces complex across the ZoA. Their sample peaked around the 
distance of this large-scale structure, namely 4250 -- 8000\kms. For comparison the detection rate of 
their nearer, optically visible, and partly classifiable into rough morphological type, is 53\% 
for the 369 galaxy candidates targeted for observation, compared to our 35\% of 2MASX selected Arecibo 
galaxies without previous redshift information. Their average noise was $1.3\pm0.5$mJy, comparable to 
our observations.

Both (our work and Pantoja's) are considerably deeper than the ALFA precursor observations of 
5-6 mJy rms sensitivity. Then again, the multibeam ALFA blind survey observations are powerful 
in that they do not require any previous identification of optical or NIR counterparts. They 
will therefore be much more efficient for nearby gas rich dwarfs at the highest extinction levels. 
In that sense, the Arecibo data presented here are complementary to the ALFA survey, at least for 
the Galactic anti-centre part of the ZoA visible from Arecibo (with relatively low extinction and 
star density).

The \HI\ mass distribution is without further surprises. The Arecibo data find on average more 
\HI-massive galaxies compared to \nan, which is the effect of the galaxies being more distant on 
average. The overall \HI-mass distribution is quite similar to the (as yet mostly unpublished) deep 
Parkes \HI\ ZoA surveys (RCKK for the ZoA team; see also Donley et al. 2005 for the northern extension; 
Shafi 2008 for the Galactic Bulge extension) with the majority of galaxies lying in the range of 
9 -- 10.5 log\MHI\ (\Msun) with a few outliers down to lower masses of a few times $10^7$\Msun, and 
two above that range ($ > 3 \cdot 10^{10}$\Msun). Both the faintest and most massive galaxy are 
peculiar and discussed in further detail in Sect. 4.3.

%
The $K_s$-band luminosity distribution is not dissimilar to the \HI-mass distribution except for an 
overall shift of one dex in the logarithmic solar units scale. This implies that the overall \HI-mass to 
$K_S$-band light ratio  has a mean of about 0.1, which corresponds closely to other surveys.
The more nearby \nan\ objects have a slightly lower \MHILK\ compared to the Arecibo observations, due to the 
slightly different selection criteria, with the more distant, small and compact Arecibo galaxies more 
likely to be high surface brightness massive spirals.

%
The estimated dynamical masses are smaller than the combined stellar and gas masses for many of the galaxies. 
There are several biases contributing to this problem that stem from the location of these galaxies in 
the ZoA. Extinction effects are probably not the main cause of this problem: although the isophotal radii 
of the galaxies are underestimated, so too are the total $K_s$-band luminosities. In fact, by pushing the 
effective isophotal surface brightness fit to a higher level (like for a less obscured  galaxy), the axis 
ratio would be  measured closer to the bulge and likely to be rounder than if it were measured farther out 
in the disc (e.g. Cameron 1990). After correcting for inclination effects, this would generally lead to  
an overestimate of the dynamical mass.

The more likely explanation therefore is that the stellar confusion is the culprit. Unidentified faint 
stars, which are plentiful at these low latitudes, can add to the $K_s$-band light and distort the shape 
of the galaxy isophotes. The latter effect would diminish the inclination correction to the rotation 
velocities, which has a strong effect on our dynamical mass indicator.

\subsection{Extinction model tests} 
\begin{figure} 
\centering
\includegraphics[width=9cm] {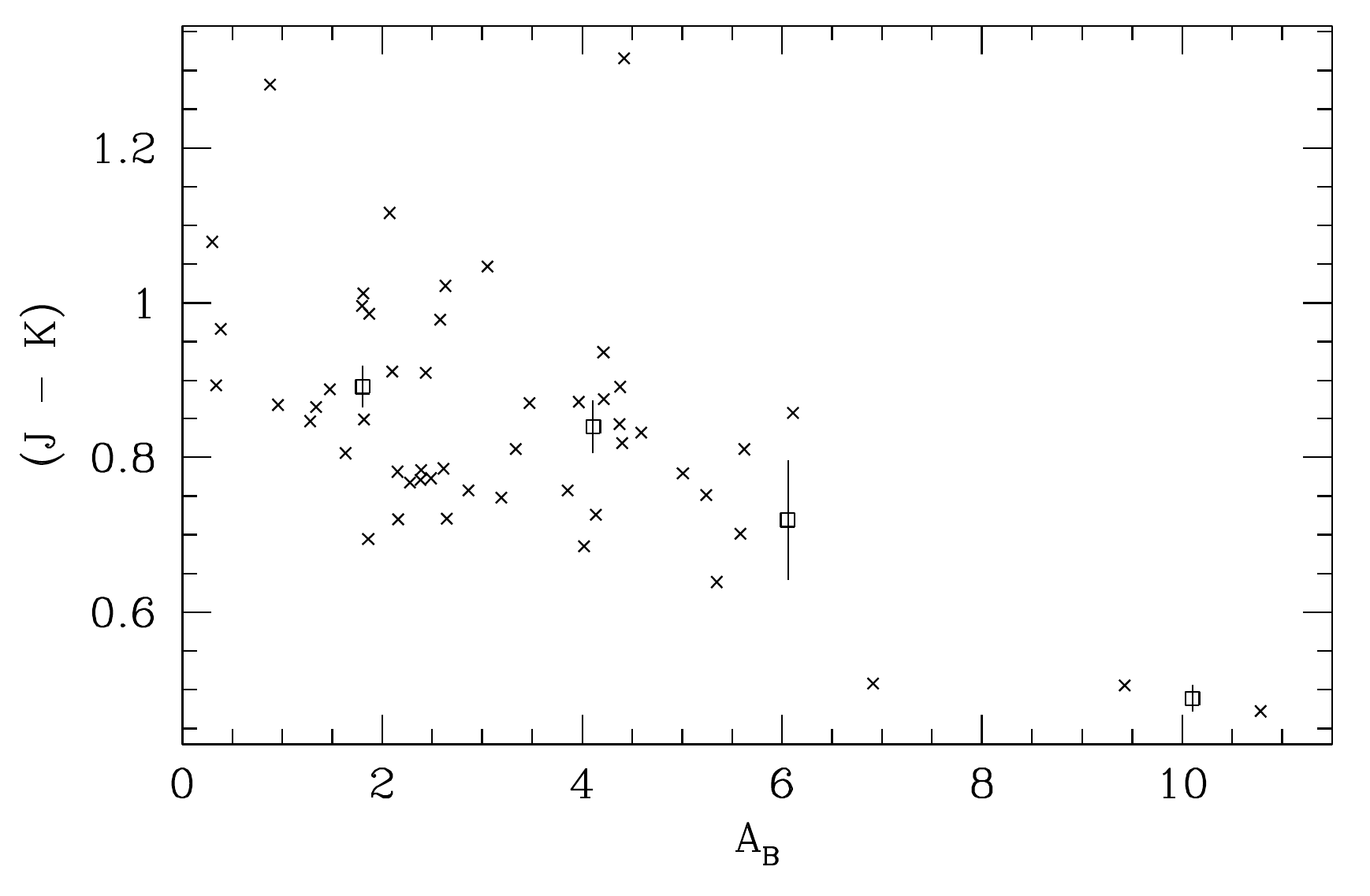}
\caption {Extinction-corrected $(J-K)^0$ colours of the \HI\ detected
galaxies as a function of the Galactic foreground $B$-band extinction correction factor in their 
direction, $A_B$, as estimated by Schlegel et al. (1998). }
\label{extinction}%
\end{figure}

The galaxies detected in this study mostly have \HI\ masses in the range of $10^9$ to $10^{10}$ \Msun\ 
(see Fig.~\ref{histo}), typical of other \HI\ studies (e.g., Roberts \& Haynes 1994), and the galaxies 
generally show no obvious trends as a function of redshift or local extinction. However, after applying 
extinction corrections at $J$ and $K_s$, and the minor corrections to the $J-K$ colours for redshift 
(Table 2 and 5, column 5), 
we still find that the galaxies grow successively redder in regions of higher estimated extinction -- 
as suggested already in section 5.2 (Fig.~\ref{colours}). Figure~\ref{extinction} shows a plot of corrected 
$(J-K)^o$ colour as a function of the value of $A_B$ estimated by Schlegel et al. (1998). This effect 
has been noted for other ZoA galaxies (Nagayama et al. 2004; Schr\"oder et al. 2007; Tagg 2008; Cluver 2009), 
who find overestimates of $f(A_{\lambda}^{real} / A_{\lambda}^{obs})=0.67, 0.87, 0.75$, and $0.84$, 
respectively, of the values implied by Schlegel et al. (1998). The thorough analysis by Schr\"oder et al. 
(2007) derived by optimising this factor in fitting all three DENIS NIR colours $IHK$ simultaneously 
to medium to highly obscured ZoA galaxies finds that reducing the extinction corrections by about 15\% 
introduces less bias as a function of $A_B$.

\begin{figure} 
\centering
\includegraphics[width=9cm] {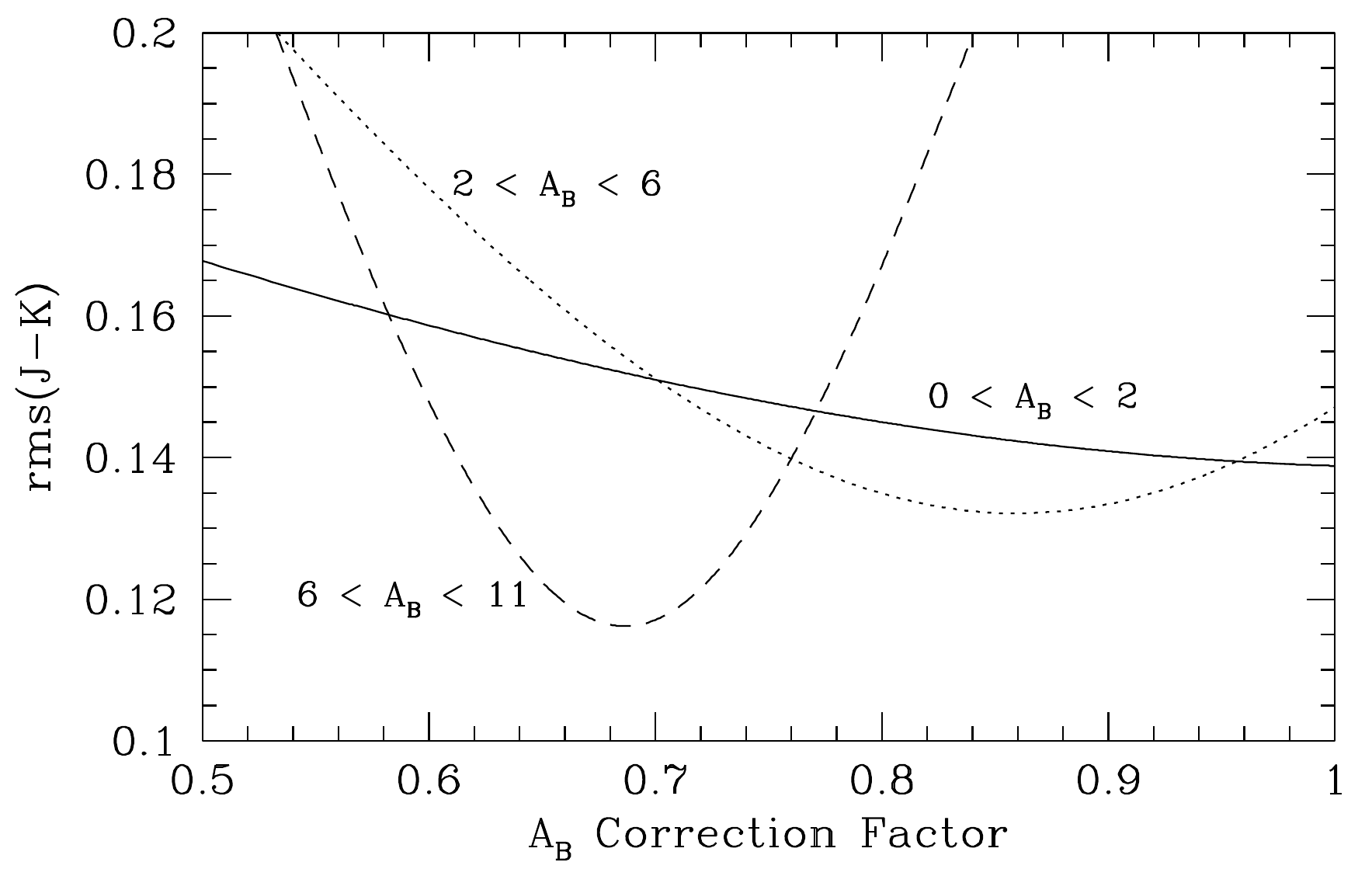}
\caption {Results of the search for the adjustments to the Galactic
$B$-band extinction correction factor, $A_B$, as estimated by
Schlegel et al. (1998), that would give the least scatter about the
expected mean $J-K$ colour of the \HI\ detected galaxies. The galaxies were divided into 3 bins, 
with $A_B$ extinction correction factors between 0-2 mag, 2-6 mag, and 6-12 mag. }
\label{chi}%
\end{figure}

Since our galaxies should intrinsically have typical $J-K$ colours of about 0\fm9, we performed the 
following test. We selected the galaxies with $A_B$ extinctions in the ranges of 0-2, 2-6, and 6-12 mag, 
respectively, and searched for the value of the adjustment to the Schlegel et al. values that would give 
the least scatter about the expected mean colour. The results of this test are shown in Fig.~\ref{chi} 
where the curves show that the scatter of the colour is minimized at approximately the Schlegel et al. (1998) 
extinction values for moderate extinction -- which generally are at $|b| \gsim$ 5\dgr, where the Schlegel et al. 
values are indeed said to still be valid, whereas in regions of moderate extinction this reduces to 86\% of 
the Schlegel et al. values, and to about 69\% for our most heavily-extincted galaxies. This falls well within 
the range of other ZoA studies. As previous studies were mostly restricted to the southern sky, it is reassuring 
to notice that this effect for our mostly northern sample seems to be of the same order in the southern sky.

This trend suggests that either the Schlegel et al. (1998) extinctions are overestimated in the most-highly 
extincted regions or that the relative infrared extinction is less in these regions.

\subsection{Indications of large-scale structures connections}  
\begin{figure*} 
\centering
\includegraphics[width=18cm] {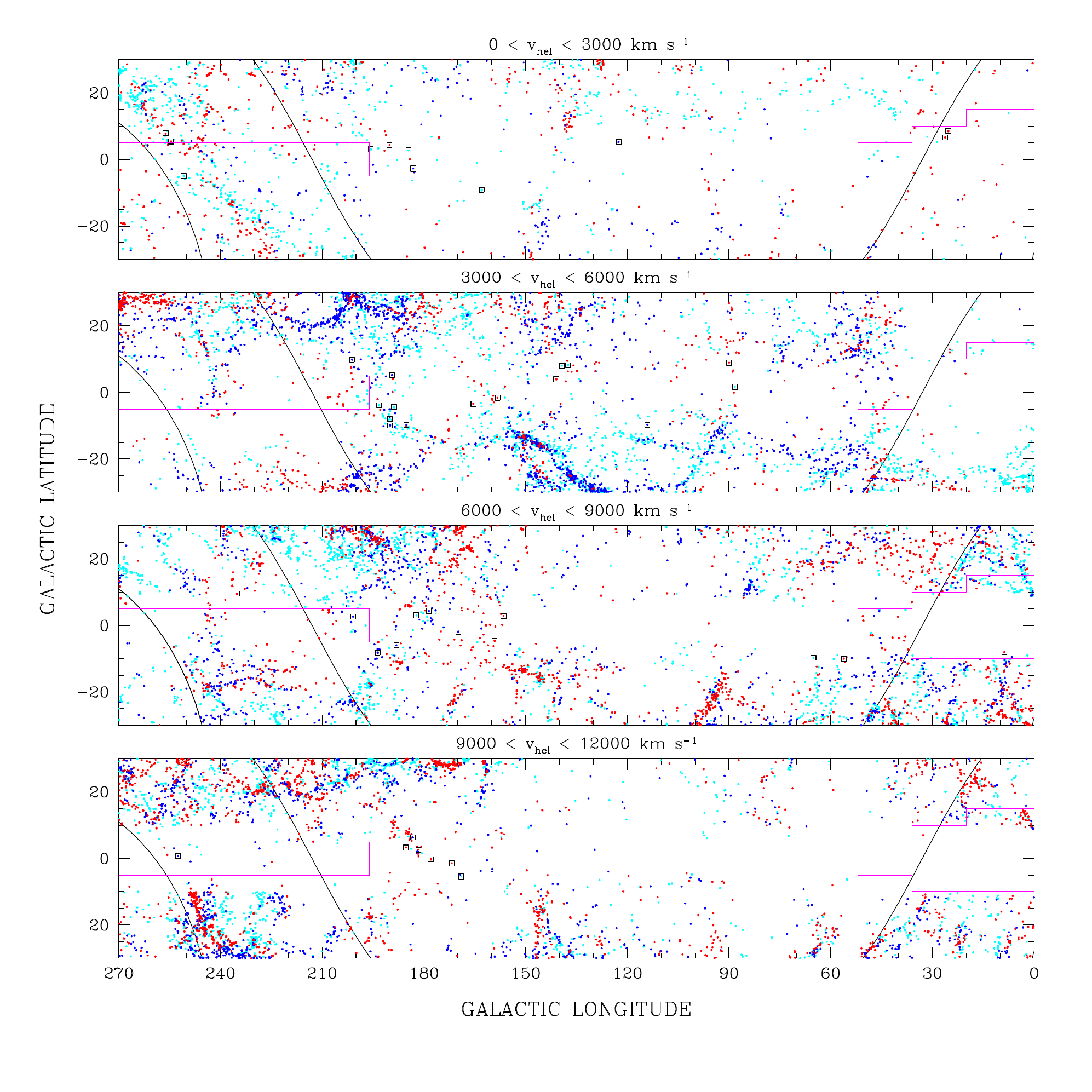}
\caption {Plots showing our \HI\ detections (black squares) superimposed on known large-scale 
structures (as available in HyperLeda) in four radial velocity intervals (0-3000, 3000-6000, 
6000-9000 and 9000-12000 \kms) in Galactic coordinates. Within each 3000 \kms\ slice in radial 
velocity, the galaxies in the nearest 1000 \kms\ wide bin are shown in red, the middle 1000 \kms\ 
in dark blue, and the furthest 1000 \kms\ in light blue. The dark lines indicate declinations of 
$\delta=12$\dgr\ as well as $\delta=-40$\dgr\, the southern  declination limits of, respectively, 
our Arecibo and \nan\ samples. The outlined boxes denote the deep \HI\ surveys undertaken with the 
Parkes MultiBeam Receiver. Note that overlap with these surveys was minimal.}
\label{LSS}%
\end{figure*}

Although this is a pilot project, and the number of newly measured redshifts of obscured galaxies is
relatively small ($N=51$), we nevertheless had a look at their distribution in redshift space to see 
what kind of structures they trace across the ZoA or what new large-scale features they might hint at. 
The locations of these new \HI\ detections are shown in Fig.~\ref{LSS}. Their positions on the sky are 
plotted as square symbols in four radial velocity slices (or shells) of 3000 \kms\ width for the velocity 
range 0 -– 12,000\kms\ (the 3 even higher velocity detections are not displayed). They are superimposed 
on the distribution of galaxies with previously measured redshifts, as obtained from HyperLeda. The colour 
coding refers to the different redshift ranges within a slice, with red marking the nearest, dark blue 
the middle and the fainter cyan the most distant $\Delta V=1000$\kms\ interval per slice. The black 
lines demarcate the southern \nan\ (-40\dgr) and Arecibo (0\dgr) observable declination limits. 

The magenta boxes outline the survey coverage of the various deep ZoA \HI-surveys (rms=6 mJy) centred 
on the southern Galactic Plane undertaken with the Multi-Beam Receiver of the Parkes 64m radio telescope 
(see Kraan-Korteweg et al. 2008 for survey details). The displayed data points also include the more shallow 
HIPASS survey results (Meyer et al. 2004; Wong et al. 2006). Note again that there is little overlap between 
our survey and this southern one, partly because of our selection of northern telescopes, and partly because 
2MASS does not uncover galaxies in the high star-density around the wider Galactic Bulge area (Fig.~9 in 
Kraan-Korteweg 2005). 
 
Overall, the new detections are mostly seen to follow the filamentary
structures. We will comment on the detection slice by slice by comparing them to known structures, 
the northern, very shallow (40mJy, $V < 4000$\kms) \HI\ Dwingeloo Obscured Galaxy Survey (DOGS; e.g. 
Henning et al. 2000a), the 2MASX Redshift survey (2MRS) Wiener Filter density field reconstructions 
(2M-WF; Erdogdu et al. 2006b), as well as the most recent 2MASX photometric redshift slices (2MX-LSS; 
Jarrett (2008) -- see http://web.ipac.caltech.edu/staff/jarrett/lss/index.html).

{\sl Top panel (0 \kms$<$$V_{hel}$$<$3000 \kms):} The galaxies around $\ell$$\sim$240\dgr\ clearly 
form part of the well-established nearby Puppis filament, whereas the detections clumped around 
180\dgr--200\dgr\ are more of a mystery. They seem to lie in a fairly underdense region. Interestingly 
enough, the 2M-WF reconstruction finds a clear unknown overdensity (called C5) in this region which 
is also notable in the respective 2M-LSS slice. This region and overdensity might be worthwhile 
pursuing further. No new galaxies were detected at the Super-Galactic Plane (SGP) crossing (at 
$\sim140$\dgr) though some are visible in the next panel. The two galaxies at about 30\dgr\ must 
form part of a nearby filamentary structure identified for the first time in the Parkes MultiBeam 
Galactic Bulge Extension survey (Kraan-Korteweg et al. 2008; Shafi 2008) that protrudes into the 
Local Void.

{\sl Second panel (3000 \kms$<$$V_{hel}$$<$6000 \kms):} The two detections at both $\sim 160\dgr$ 
and $\sim 90\dgr$ are clearly connected with the Perseus-Pisces (PP) complex. The connection 
across the ZoA leading from Perseus to A569 is very strong in the 2M-WF reconstructions as well 
as evident in the 2M-SS slice, more clearly so in the heavily smoothed display. The two other 
galaxies belong to the south-western spur of the PP complex, which was also found by 2M-WF and 
2M-LSS. Both the detections at $\sim 190\dgr$ slightly below the plane) and $\sim140\dgr$ (above 
the plane) show a prominent signal in 2M-WF (marked as OR for Orion and CAM for Cameleopardis there). 
While visible in 2M-LSS it is less obvious there, though more so in the smoothed version.

{\sl Third panel (6000 \kms$<$$V_{hel}$$<$9000 \kms):} Two clouds of detections can be attributed 
to known, or rather suspected, structures. The two galaxies at $\sim 190\dgr$ (below the plane) 
also seem part of Orion (OR), as this has an even stronger signal in the reconstructed density 
field in 2M-WF for this redshift range. The other galaxies seem to follow the main PP-chain. 

{\sl Bottom panel (9000 \kms$<$$V_{hel}$$<$12,000 \kms):} Six of the 7 detections are remarkably 
aligned and seem to suggest some kind of filament -- or the far end of the sheet-like PP-chain? 
Such a feature is not recovered in the 2M-WF, however, though it is evident in the 2M-LSS slices. 
It might be worthwhile to observe more galaxies in this velocity  range to verify whether this 
truly is a previously unknown filament.

\section{Conclusions}  
To complement ongoing ``all-sky'' redshift surveys to map extragalactic large-scale structures 
in the nearby Universe and improve our understanding of its dynamics and observed flow fields, 
we undertook a pilot project to obtain \HI\ observations of about 200 optically obscured or invisible 
galaxy candidates behind the Milky Way ($|b| <10\dgr$ ), the Zone of Avoidance strip generally avoided 
in such surveys. Likely galaxy sources were extracted in 2000-2002 from the 2MASS Extended Source 
Catalogue (2MASX), then a database under construction. This near-infrared (NIR) catologue penetrates the 
ZoA to considerably lower extinction levels than optical ones. For this we used both the \nan\ and 
Arecibo radio telescopes.

Apart from excluding extremely blue objects, no further selection criteria were applied, as 
near-infrared galaxy colours show hardly any dependence on morphology. Furthermore, the NIR galaxy 
colours are affected quite strongly by the varying Galactic dust column density through which the 
galaxies are viewed.
 
The overall detection rate of the 185 observed 2MASX galaxies whose spectra were not affected by
nearby continuum sources is quite low: 24\% and 35\% for the \nan\ and Arecibo samples respectively 
This detection rate is lower than for \HI-follow-ups of optically selected galaxies, even in the ZoA. 
For instance, a similar \HI\ survey of optically selected ZoA galaxies (Kraan-Korteweg et al. 2002) 
reached a 44\% detection rate (though a pre-selection on morphological type favoring spiral galaxies 
was made).

Despite this relatively low detection rate, it should be noted that other means of obtaining redshifts 
(optical spectroscopy) for galaxies hidden by the Milky Way remain extremely difficult due to
their reduced surface brightness. \HI\ observations of galaxy candidates still remain the most efficient 
tool in mapping large-scale structures across the ZoA.

In addition, this pilot survey taught us that  a significant number of the non-detections could 
have easily been excluded from the observing list by \\
-- (a) examination of the composite $JHK$ images of the {\sl extinction-corrected} brightest 
sources (now easily available), in combination with  the higher-resolution optical SDSS images (when available 
 - which is rarely the case for the present sample). Most of the 
extended objects with $K^o \ga 7\fm0$ or $7\fm5$ could readily be dismissed as galaxy candidates in 
that way (see Fig.~5 and Sect. 5.1) \\
-- (b) considering {\sl extinction-corrected} colour limits. All objects that were bluer than 
$(J-K)^0 < 0\fm5$ and $(H-K)^0 < 0\fm0$ (Fig.~6) were found to be Galactic objects, mostly \HII\ regions 
or filamentary structures associated with Galactic objects (Sect.~5.5).

 Based on these two criteria the total number of \nan\ sources observed would have been reduced by 
15\%, to 99 instead of 116: 11 on their images alone, 6 just on their colour, and 7 on both. None of these 
17 objects were detected in \HI.
The smaller and fainter Arecibo sources are not really adequate for rejection through NIR image examination, 
as the largest and brightest ones in the area had already been observed in \HI\ by Pantoja et al. (1994, 1997). 
Only two of them would have been rejected based on their colours.

The Schlegel et al. (1998) Galactic extinction values serve as a good first proxy for these tests, 
even though they are not calibrated at the lowest Galactic latitudes ($|b| \le 5\dgr$). An extinction 
overestimate will have a minimal effect on the verification procedures, or even on the extinction-corrected 
colours delimitations as the selective reddening will reduce the colours by only a relatively low amount 
of $0.12A_B$ and $0.05A_B$ for $J-K$ and $H-K$ respectively.

The NIR colours of the detected galaxies were actually used to assess the accuracy of the DIRBE 
extinction values at low latitudes if taken at face value.  The results confirm that the values from
Schlegel et al. (1998) are valid for latitudes above $|b| \ga 5\dgr$, whereas in regions of moderate 
extinction this reduces to 86\%, and to 69\% for our most heavily-extincted galaxies. 
It is reassuring that these values (from our mostly northern sky sample) seems in good agreement with 
the previous mostly southern sky derivations (e.g., Schroeder et al. 2007 and references therein).

Overall, the properties of the detected galaxies match those of other surveys. The sample is too sparse 
yet to give an improved insight into suspected or unknown large-scale structures behind the Milky
Way. However, the detections already indicate (see Fig.~10) that a further probing of the galaxy 
distribution will quantify filaments, clusters and also voids in this part of the ZoA.

Hence our \HI\ detection rate of 2MASS-selected ZoA galaxies can be significantly improved, 
if the above mentioned image examinations and extinction-corrected limits are employed. 
Such a systematic survey would actually be a worthy pursuit, as it would be complementary to the 
ongoing ``blind'' deep ZoA \HI\ Parkes multi-beam survey (at $|b| \le 5\dgr$; of similar velocity 
coverage though slightly lower sensitivity than the \nan\ pilot project), if done for the ZoA 
in the latitude range of $5\dgr \le |b| \le 10\dgr$ accessible to \nan, and for the northern ZoA
at $|b| \le 10\dgr$ in the areas not covered by the Arecibo ALFA surveys ($\delta > 35\dgr$). 
We intend to pursue such a survey at \nan.

\acknowledgements{
This publication makes use of data products from the Two Micron All
Sky Survey, which is a joint project of the University of
Massachusetts and the Infrared Processing and Analysis Center, funded
by the National Aeronautics and Space Administration and the National
Science Foundation. We also wish to thank the Arecibo Observatory
which is part of the National Astronomy and Ionosphere Center, which
is operated by Cornell University under a cooperative agreement with
the National Science Foundation. This research has made use of the
HyperLeda database (http://leda.univ-lyon1.fr), the NASA/IPAC
Extragalactic Database (NED) which is operated by the Jet Propulsion
Laboratory, California Institute of Technology, under contract with
the National Aeronautics and Space Administration, and the Aladin
database, operated at CDS, Strasbourg, France. We acknowledge financial 
support from CNRS/NSF collaboration grant No.10637 and from the ASTE of CNRS/INSU. 
RCKK wishes to thank the South African National Research Foundation for support. 
  }

\onecolumn

\newpage
\clearpage 

%

\begin{landscape}
\begin{table*}  
\caption{\nan: observational data}
\centering
{\footnotesize
\begin{tabular}{crrlrlrrrrlrrrrrrrrrr}
\hline\hline             
No & 2MASX J          & PGC No  & Other Name        & $K_{20}$ & $J$-$K$& $H$-$K$ &$r_{K20}$& $b/a$  &$B_{T_C}$&$D_{25}$&$I_{HI}$&$V_{50}$&$W_{50}$&$W_{20}$&$rms$&$S_{max}$ \\
    &                  &          &                   & mag      & mag  &  mag &  $''$   &     
& mag   & $'$      & \JYKMS\ & \KMS\ & \KMS\ & \KMS\ & mJy & mJy  \\
(1)  &  (2) & (3) & (4) & (5) & (6) & (7) & (8) & (9) & (10) & (11) & (12) & (13) & (14) & (15) & (16) & (17) \\
\hline
N01 & 00475430+6807433 &   137211 & ZOAG 122.60+05.26    & 10.86 & 1.41 & 0.49 & 37.00 &  0.38 &     * &  0.2 & 11.7$\pm$0.6 &   761$\pm$1 & 338 & 351 &  3.93 &  58 \\ 
N02 & 01203021+6525055 &   137634 & ZOAG 125.95+02.72    &  9.96 & 1.43 & 0.45 & 40.10 &  0.58 &     * &  0.4 & 15.8$\pm$0.9 &  4150$\pm$3 & 419 & 447 &  5.02 &  80 \\ 
N03 & 03261404+6635372 &  2797282 & ZOAG 137.47+08.19    &  8.98 & 1.23 & 0.38 & 44.00 &  0.74 &     * &  1.8 &  1.0$\pm$0.3 &  5460$\pm$2 & 101 & 111 &  3.15 &  20 \\ 
N04 & 03302327+6110135 &   168353 & ZOAG 140.93+03.98    &  9.39 & 1.49 & 0.43 & 42.90 &  0.56 & 12.12 &  0.7 &  3.0$\pm$0.4 & 3678$\pm$16 & 231 & 281 &  3.04 &  14 \\ 
N05 & 03394709+6528486 &  2677859 & ZOA 139.25+08.09     & 10.15 & 1.43 & 0.40 & 37.70 &  0.52 & 11.89 &  0.7 & 10.2$\pm$0.9 &  5219$\pm$1 & 437 & 446 &  5.02 &  52 \\ 
N06 & 04212943+3656572 &        * & *                    &  9.31 & 1.26 & 0.35 & 34.80 &  0.88 &     * &    * &  1.1$\pm$0.3 &  2435$\pm$5 &  80 & 102 &  2.36 &  16 \\ 
N07 & 04244617+4244494 &   101333 & ZOAG 159.16-04.63    & 10.42 & 1.16 & 0.35 & 43.20 &  0.20 & 11.58 &  1.5 &  3.3$\pm$0.4 &  6018$\pm$3 & 506 & 521 &  1.98 &  15 \\ 
N08 & 04333811+4530061 &    15526 & ZOAG 158.26-01.58    &  9.66 & 1.54 & 0.48 & 52.90 &  0.36 &     * &  0.7 &  7.2$\pm$0.4 &  3858$\pm$3 & 320 & 339 &  2.48 &  27 \\ 
N09 & 04464159+4943063 &  2798769 & ZOAG 156.57+02.86    & 11.00 & 1.29 & 0.41 & 34.50 &  0.58 &     * &  0.9 &  3.6$\pm$0.5 &  6582$\pm$3 & 367 & 381 &  2.83 &  18 \\ 
N10 & 04514426+3856227 &   168835 & ZOAG 165.41-03.37    & 10.56 & 1.43 & 0.42 & 38.30 &  0.28 &     * &  0.9 &  8.2$\pm$0.6 &  3919$\pm$3 & 357 & 374 &  3.56 &  36 \\ 
N11 & 07492337-3542214 &  2807067 & HIZSS 25             & 10.69 & 1.64 & 0.46 & 38.20 &  0.68 &     * &  0.8 & 15.1$\pm$0.2 &  2863$\pm$1 &  47 &  69 &  3.35 & 300 \\ 
N12 & 08080461-1452387 &    79913 & FGC 717              & 10.69 & 1.17 & 0.36 & 34.90 &  0.24 & 13.27 &  1.9 &  5.8$\pm$0.5 &  6575$\pm$3 & 463 & 481 &  2.53 &  22 \\ 
N13 & 08170147-3410277 &        * & 2MFGC 6552           & 10.74 & 1.46 & 0.50 & 39.20 &  0.16 &     * &    * &  9.2$\pm$0.7 & 10373$\pm$5 & 622 & 651 &  2.94 &  27 \\ 
N13 S & (0816586-3420024) &     * &                      &       &      &      &       &       &       &      &  6.4$\pm$0.6 & 10380$\pm$5  & (207) & 592 & 2.74 & 22 \\
N14 & 08410265-3303095 &    24405 & ESO 371-5            & 11.24 & 1.02 & 0.17 & 40.70 &  0.20 & 13.16 &  1.1 &  1.2$\pm$0.3 & 1901$\pm$16 & 185 & 240 &  1.99 &  10 \\ 
$[$N15$]$ & 08544150-3248590 & 25014 & ESO 371-027       &     * &    * &    * &     * & (0.4) & 15.46 &  0.6 & $>$2.2       &  1302       &  73 &  98 &  3.04 &  36 \\ 
N16 & 16183236-3723459 &   623805 & *                    &  9.82 & 1.34 & 0.43 & 39.40 &  0.88 &     * &  0.3 &  4.5$\pm$0.8 & 4584$\pm$12 & 156 & 206 &  6.22 &  38 \\ 
N17 & 18071027-0249517 &  2801967 & HIPASS 1807-02       & 10.27 & 1.38 & 0.40 & 36.50 &  0.84 &     * &  1.1 & 31.3$\pm$0.1 &  1767$\pm$1 & 143 & 165 &  1.27 & 280 \\ 
N18 & 18153013-0253481 &   166531 & *                    &  9.05 & 1.69 & 0.46 & 69.00 &  0.57 &     * &  0.3 &  9.2$\pm$0.3 &  1786$\pm$1 & 313 & 329 &  1.99 &  42 \\ 
N19 & 18360975-2505397 &   207189 & CGMW 4-1770          &  9.78 & 1.29 & 0.38 & 44.60 &  0.38 &     * &  1.3 & 10.5$\pm$0.6 &  6572$\pm$4 & 608 & 635 &  2.91 &  33 \\
N20 & 20272088+5357579 &  2455941 & *                    &  9.99 & 1.10 & 0.30 & 34.30 &  0.52 & 13.71 &  0.8 &  2.7$\pm$0.3 &  3281$\pm$2 & 217 & 228 &  2.02 &  21 \\ 
N21 & 20572285+4808542 &        * & *                    & 10.42 & 1.86 & 0.58 & 41.00 &  0.26 &     * &    * &  6.5$\pm$0.7 &  5697$\pm$7 & 417 & 441 &  4.04 &  22 \\ 
N22 & 21135161+4255323 &        * & 2MFGC 16082          & 13.16 & 1.28 & 0.23 & 16.5  &  0.28 &     * &    * &  3.9$\pm$  &  4437$\pm$  & 277 & 301 & 3.27 & 23 \\
N23 & 23535349+5210172 &    72790 & UGC 12836            & 11.35 & 1.05 & 0.24 & 29.80 &  0.34 & 13.34 &  1.0 &  4.7$\pm$0.5 &  4724$\pm$2 & 336 & 346 &  3.04 &  25 \\ 
\hline
\multicolumn{17}{l}{{\bf Notes}: no. N15: galaxy ESO 371-27 was detected in the beam centred on the undetected target, ESO 371-26 (see Sect. 4.1).}  \\
\end{tabular}
}
\vfill
\end{table*}
\end{landscape}

\newpage
\clearpage 

\begin{landscape}
\begin{table*}  
\caption{\nan: derived data}
\centering
{\footnotesize
\begin{tabular}{crrrrlrrrrrrrrrrrrrrrr}
\hline\hline             
No. &   $l$  &    $b$ & $A_B$ &  $D$  &$k_{J-K}$& $R_{K20}$ &log($M_{HI}$)&log($L_K$)&log$({M_{HI}\over L_K})$&
log($L_B$)& {\bf log$({M_{HI}\over L_B})$} &$v_{rot}$&log($M_{dyn}$)&log$({M_{baryon}\over M_{dyn}})$   \\
    &   deg  &    deg & mag & Mpc   & mag  & kpc  & \Msun\ & \LsunK\ &   & \LsunB\ &       & \KMS\ & \Msun\ &  \\
(1)  &  (2)  & (3)    & (4) & (5)   & (6)  & (7)  & (8)   & (9)   & (10)  & (11)  & {\bf (12)}  & (13)  & (14)  & (15) \\
\hline
N01 & 122.60 &  5.26 & 5.01 &  15.0 & 0.01 &  2.7 &  8.79 &  9.50 & -0.71 &     * &     * & 182.7 & 10.32 & -0.79 \\  
N02 & 125.95 &  2.72 & 5.58 &  60.9 & 0.04 & 11.8 & 10.14 & 11.10 & -0.96 &     * &     * & 257.2 & 11.26 & -0.18 \\ 
N03 & 137.46 &  8.19 & 4.02 &  78.0 & 0.05 & 16.6 &  9.17 & 11.65 & -2.48 &     * &     * &  75.1 & 10.33 &  1.23 \\  
N04 & 140.93 &  3.98 & 4.21 &  52.6 & 0.03 & 10.9 &  9.29 & 11.15 & -1.86 & 10.79 & -1.50 & 139.4 & 10.69 &  0.38 \\ 
N05 & 139.25 &  8.00 & 4.37 &  74.4 & 0.04 & 13.6 & 10.12 & 11.16 & -1.03 & 11.18 & -1.06 & 255.8 & 11.31 & -0.18 \\ 
N06 & 162.89 & -9.13 & 4.14 &  34.3 & 0.02 &  5.8 &  8.47 & 10.81 & -2.34 &     * &     * &  84.2 &  9.97 &  0.75 \\  
N07 & 159.16 & -4.63 & 2.62 &  84.3 & 0.05 & 17.7 &  9.74 & 11.10 & -1.35 & 11.41 & -1.67 & 258.2 & 11.43 & -0.39 \\  
N08 & 158.25 & -1.58 & 5.62 &  53.2 & 0.03 & 13.6 &  9.68 & 11.10 & -1.42 &     * &     * & 171.5 & 10.97 &  0.06 \\  
N09 & 156.56 &  2.86 & 3.85 &  92.4 & 0.05 & 15.5 &  9.87 & 10.99 & -1.12 &     * &     * & 225.2 & 11.26 & -0.31 \\  
N10 & 165.41 & -3.37 & 4.22 &  53.3 & 0.03 &  9.9 &  9.74 & 10.70 & -0.96 &     * &     * & 185.9 & 10.90 & -0.22 \\ 
N11 & 250.83 & -4.89 & 6.11 &  41.1 & 0.02 &  7.6 &  9.78 & 10.48 & -0.70 &     * &     * &  32.0 &  9.26 &  1.26 \\  
N12 & 235.16 &  9.51 & 0.30 &  92.7 & 0.05 & 15.7 & 10.07 & 10.99 & -0.92 & 10.82 & -0.75 & 238.5 & 11.31 & -0.33 \\  
N13 & 252.54 &  0.72 & 2.07 & 147.9 & 0.09 & 28.1 & 10.68 & 11.44 & -0.76 &     * &     * & 315.0 & 11.81 & -0.35 \\  
N14 & 254.56 &  5.43 & 1.28 &  23.9 & 0.01 &  4.7 &  8.23 &  9.63 & -1.40 &  9.68 & -1.45 &  94.4 &  9.98 & -0.41 \\  
$[$N15$]$ & 256.17 & 7.82 & 1.36 & 14.3 & 0.01 & * & 8.37 &     * &     * &  8.32 &  0.05 &     * &     * \\  
N16 & 342.15 &  9.24 & 5.34 &  63.5 & 0.04 & 12.1 &  9.64 & 11.18 & -1.55 &     * &     * & 164.2 & 10.88 &  0.23 \\  
N17 &  25.30 &  8.52 & 6.91 &  23.6 & 0.01 &  4.2 &  9.62 & 10.20 & -0.58 &     * &     * & 131.8 & 10.22 &  0.05 \\  
N18 &  26.22 &  6.64 & 9.43 &  23.9 & 0.01 &  8.0 &  9.09 & 10.78 & -1.69 &     * &     * & 190.5 & 10.82 & -0.12 \\  
N19 &   8.79 & -8.03 & 1.81 &  90.3 & 0.05 & 19.5 & 10.31 & 11.39 & -1.08 &     * &     * & 328.6 & 11.69 & -0.33 \\  
N20 &  89.96 &  8.97 & 1.47 &  49.4 & 0.03 &  8.2 &  9.20 & 10.77 & -1.57 & 10.10 & -0.90 & 127.0 & 10.48 &  0.22 \\  
N21 &  88.24 &  1.69 &10.79 &  83.1 & 0.05 & 16.5 & 10.03 & 11.36 & -1.34 &     * &     * & 215.9 & 11.25 &  0.05 \\  
N22 &  86.29 & -3.95 & 3.64 &  65.5 & 0.04 &  5.2 &  9.60 &  9.82 & -0.22 &     * &     * & 144.3 & 10.40 & -0.36 \\  
N23 & 114.04 & -9.70 & 1.64 &  70.3 & 0.04 & 10.2 &  9.74 & 10.53 & -0.79 & 10.55 & -0.81 & 178.6 & 10.87 & -0.33 \\ 
\hline
\multicolumn{15}{l}{{\bf Notes}: no. N15: galaxy ESO 371-27 was detected in the beam centred on the undetected target, ESO 371-26 (see Sect. 4.1).}  \\
\end{tabular}
}
\vfill
\end{table*}
\end{landscape}

\newpage
\clearpage

\begin{table}  
\caption{\nan: non-detections}
 \centering
{\footnotesize           
\begin{tabular}{lrlrrrrrrrl}
\hline\hline
2MASX J              &  PGC    & others                  & $l$   & $b$   & $A_B$ & $K_{20}$ & $J$-$K$ & $H$-$K$ & rms   & V$_{opt}$ \\ 
                     &         &                         & (deg) & (deg) & (mag) & (mag)    & (mag)   & (mag)   & (mJy) &  (\kms)    \\
\hline
00094437+6520217     &         &                         & 118.60 &  2.82 & 13.04 & 10.06 & 1.53 & 0.63 &     3.4  &            \\  
00103893+5325040     & 2440513 &                         & 116.80 & -8.97 &  1.12 & 10.28 & 1.10 & 0.32 &     2.9  &            \\  
00234412+5538298*    &         &                         & 119.00 & -7.02 &  1.82 & 13.33 & 1.40 & 0.51 &     3.2  &            \\  
00253292+6821442     & 136991  &  ZOAG 120.54+05.61      & 120.54 &  5.61 &  4.22 & 10.04 & 1.37 & 0.41 &     9.2  &            \\  
00584741+5627495     &         &                         & 123.95 & -6.39 &  2.53 & 10.48 & 1.33 & 0.62 &     3.0  &            \\  
01010771+6254430     &         &                         & 124.03 &  0.06 & 11.10 & 10.82 & 1.99 & 0.93 &     3.6  &            \\  
01202398+5801226*    & 4831    &  ZOAG 126.77-04.63      & 126.77 & -4.63 &  2.03 &  9.65 & 1.24 & 0.41 &     2.5  &            \\  
01281012+6313517     &         &                         & 127.05 &  0.65 &  6.75 & 10.30 & 1.62 & 0.50 &     4.9  &            \\  
01452714+6414338     &         &                         & 128.76 &  1.98 &  8.82 & 13.24 & 1.76 & 0.70 &     3.6  &            \\  
01465173+6618392     & 137901  &  ZOAG 128.47+04.04A     & 128.47 &  4.04 &  6.79 &  9.37 & 1.72 & 0.52 &     3.8  &            \\  
01592758+6606246     & 138017  &  ZOAG 129.76+04.14      & 129.75 &  4.14 &  4.72 &  9.73 & 1.62 & 0.48 &     7.5  &            \\  
02113142+7046204*    & 137982  &  ZOAG 129.49+08.92      & 129.49 &  8.93 &  3.81 &  8.63 & 1.33 & 0.42 &     6.7  &            \\  
02161335+7041194*    & 138040  &  ZOAG 129.89+08.97      & 129.88 &  8.97 &  4.06 &  8.97 & 1.34 & 0.38 &     3.8  &            \\  
02204463+5349396     &         &  HFLLZOA G135.98-06.77  & 135.98 & -6.77 &  0.87 & 10.97 & 1.23 & 0.37 &     4.8  &            \\  
02421337+6723309     & 2796960 &  ZOAG 133.30+06.78      & 133.30 &  6.78 &  4.72 & 10.37 & 1.36 & 0.38 &     7.0  &            \\  
02490220+6304550     & 168303  &  ZOAG 135.80+03.19      & 135.79 &  3.18 &  4.10 &  9.75 & 1.56 & 0.52 &     6.6  &            \\  
02491146+6628401     & 2797016 &  ZOAG 134.32+06.24      & 134.31 &  6.24 &  4.55 &  9.89 & 1.28 & 0.39 &     5.0  &            \\  
02503998+5316109     & 2797482 &  ZOAG 140.32-05.53      & 140.31 & -5.53 &  2.19 & 10.52 & 1.33 & 0.40 &     4.9  &            \\  
03012738+6030399     &         &                         & 138.28 &  1.57 & 25.42 & 13.40 & 1.59 & 0.58 &    11.4  &            \\  
03210915+6655186     & 2797207 &  ZOAG 136.87+08.19      & 136.86 &  8.19 &  4.60 &  8.67 & 1.33 & 0.35 &     4.1  &            \\  
03292042+6601389*    & 2682877 &  IRAS 03248+6551        & 138.05 &  7.91 &  4.84 &  9.61 & 1.47 & 0.41 &     5.9  &            \\  
03381210+6642567     & 2690279 &                         & 138.37 &  8.98 &  4.80 &  9.47 & 1.48 & 0.41 &     6.0  &            \\  
03445127+4550430     & 13736   &  MCG+08-07-011          & 151.74 & -7.10 &  1.95 & 10.43 & 1.22 & 0.36 &     8.0  &            \\  
03482003+4156074     & 13875   &  UGC 2863               & 154.70 & -9.77 &  1.49 & 10.22 & 1.12 & 0.32 &     3.8  &            \\  
04052826+4938132     & 2348913 &  ZOAG 151.97-01.98      & 151.97 & -1.97 &  7.74 & 11.02 & 1.85 & 0.94 &     3.4  &            \\  
04073579+3919305     &         &  2MFGC 3367             & 159.23 & -9.34 &  3.60 & 10.41 & 1.29 & 0.37 &     7.3  &            \\  
04210637+3704473     & 2092572 &  2MFGC 3527             & 162.75 & -9.10 &  5.01 & 10.04 & 1.44 & 0.44 &     3.0  &            \\  
04215207+3607373     & 14959   &  UGC 3021               & 163.54 & -9.65 &  2.28 &  9.19 & 1.19 & 0.37 &     2.8  &  6238$^1$  \\  
04390214+4608079     & 2276003 &  ZOAG 158.42-00.46      & 158.42 & -0.46 &  6.79 &  9.04 & 1.62 & 0.51 &     2.9  &            \\  
04391275+5820444     &         &  2MFGC 3806             & 149.28 &  7.67 &  2.28 &  9.98 & 1.24 & 0.37 &     3.5  &            \\  
04554245+4402134     & 16346   &  ZOAG 161.92+00.43      & 161.92 &  0.43 &  3.19 & 10.07 & 1.35 & 0.40 &     5.7  &            \\  
04562691+4555425     & 16370   &  ZOAG 160.53+01.71      & 160.53 &  1.71 &  4.06 & 10.21 & 1.51 & 0.44 &     4.0  &            \\  
04580915+4208498     & 16429   &  ZOAG 163.67+00.41      & 163.68 & -0.40 &  2.73 & 10.09 & 1.39 & 0.47 &     4.0  &            \\  
05085095+3749009     &         &  2MFGC 4200             & 168.36 & -1.42 &  4.55 & 11.80 & 1.80 & 0.63 &     2.6  &            \\  
05511351+3754024     & 18006   &  UGC 3367               & 172.90 &  5.61 &  2.48 &  9.90 & 1.31 & 0.39 &     3.8  &            \\  
06391085-0130277     & 85930   &  ZOAG 212.89-03.42      & 212.88 & -3.41 &  4.02 &  8.99 & 1.42 & 0.42 &     2.1  &            \\  
06434846-0108200     & 75984   &  ZOAG 213.08-02.22      & 213.08 & -2.22 & 16.39 &  8.47 & 1.91 & 1.05 &     4.8  &            \\  
06501768-0251397     & 76120   &  ZOAG 215.35-01.56      & 215.35 & -1.56 &  4.60 &  8.89 & 1.41 & 0.44 &     4.6  &            \\  
06514530-0336294     & 76161   &  ZOAG 216.18-01.58      & 216.18 & -1.57 &  5.01 &  9.57 & 1.41 & 0.42 &    11.2  &            \\  
07165014-1852167     &         &  2MFGC 5793             & 232.57 & -3.12 &  7.74 & 12.10 & 1.73 & 0.56 &    14.3  &            \\  
07300453-1833166     &         &  Galact. nebula         & 233.77 & -0.20 & 65.29 & 12.44 & 1.19 & 0.55 &     8.8  &            \\  
07300521-2044229     & 837601  &                         & 235.69 & -1.25 &  7.95 & 11.25 & 1.53 & 0.79 &     2.7  &            \\  
07300594-1832546     &         &                         & 233.77 & -0.19 & 65.29 & 12.38 & 1.36 & 1.08 &    10.6  &            \\  
07373770-2644489     & 77806   &  ZOAG 241.78-02.64      & 241.79 & -2.64 &  4.35 &  9.60 & 1.45 & 0.44 &     2.4  &            \\  
07415613-1812215     &         &                         & 234.84 &  2.43 &  2.53 &  9.53 & 0.78 & 0.75 &     4.4  &            \\  
07432302-2912591     & 78144   &  ZOAG 244.56-02.75      & 244.56 & -2.75 &  2.61 &  9.73 & 1.20 & 0.40 &     7.3  &            \\  
07451863-3231000     & 78286   &  ZOAG 247.64-04.03      & 247.64 & -4.03 &  6.67 &  8.71 & 1.56 & 0.45 &     3.4  &            \\  
07530935-2825588     & 78821   &  ZOAG 244.97+00.51      & 244.97 & -0.51 &  4.10 &  9.34 & 1.34 & 0.38 &     3.3  &            \\  
07551072-2341325     & 78984   &  ZOAG 241.14+02.32      & 241.14 &  2.32 &  1.70 &  9.47 & 1.17 & 0.33 &     3.3  &            \\  
08024707-1219041     & 22578   &  NGC 2517               & 232.27 &  9.73 &  0.41 &  8.95 & 0.95 & 0.28 &     2.9  &            \\  
08060746-2233492     & 22742   &  ESO 561-12             & 241.50 &  5.06 &  0.75 &  9.55 & 1.04 & 0.29 &     3.3  &            \\  
08123960-1603028     & 23011   &  IC 500                 & 236.77 &  9.83 &  0.29 &  9.82 & 0.97 & 0.28 &     4.4  &            \\  
08125048-2733115     & 23020   &  ESO 494-35             & 246.53 &  3.64 &  1.16 &  8.42 & 0.99 & 0.25 &     5.7  &  1047$^2$  \\  
08145966-3052011     & 23091   &  ESO 430-28             & 249.56 &  2.20 &  1.95 &  7.23 & 0.97 & 0.26 &     7.2  &            \\  
08172690-2440519     & 23234   &  ESO 494-42             & 244.70 &  6.09 &  0.66 & 10.17 & 1.04 & 0.33 &     2.2  &  1731$^2$  \\  
08442859-3132037     &         &                         & 253.80 &  6.93 &  0.99 & 12.85 & 1.27 & 0.44 &     3.4  &            \\  
08551578-3202478     & 25045   &  ESO 432-12             & 255.64 &  8.40 &  1.45 & 10.00 & 1.02 & 0.27 &     2.8  &            \\  
16380086-3605570     &         &                         & 345.81 &  7.28 &  3.35 &  9.44 & 1.26 & 0.38 &     4.8  &            \\  
16561212-2902288     &         &                         & 353.76 &  8.84 &  0.91 & 11.26 & 1.30 & 0.50 &     4.8  &            \\  
17023762-2616264     &         &                         & 356.86 &  9.38 &  1.28 &  5.84 & 0.77 & 0.16 &     3.7  &            \\  
17165337-2647470     &         &                         & 358.32 &  6.49 &  4.55 &  9.99 & 1.26 & 0.31 &     4.5  &            \\  
17454632-2228191     &         & Gal. Neb.?              &   5.54 &  3.33 &  4.31 &  7.35 & 1.43 & 0.51 &     4.4  &            \\  
17480455-2446441     &         &                         &   3.84 &  1.69 & 16.56 &  3.32 & 1.96 & 0.55 &     3.2  &            \\  
18085372-3705543     & 61461   &  ESO 394-30             & 355.32 & -8.28 &  0.70 &  9.29 & 0.91 & 0.24 &     3.8  &            \\  
\hline
\end{tabular}
}
\end{table}

\newpage
\clearpage

\begin{table}  
\addtocounter{table}{-1}
\caption{\nan: non-detections - continued.}
 \centering
{\footnotesize
\begin{tabular}{lrlrrrrrrrl}
\hline\hline
2MASX J              &  PGC    & others                  &  $l$   & $b$   & $A_B$ & $K_{20}$ & $J$-$K$ & $H$-$K$ & rms   & V$_{opt}$ \\  
                     &         &                         &  (deg) & (deg) & (mag) & (mag)    & (mag)   & (mag)   & (mJy) &  (\kms)    \\
\hline 
18130421-3740314     & 61586   &  ESO 335-IG3            & 355.19 & -9.28 &  0.62 &  9.61 & 1.04 & 0.28 &     5.9  &  7620$^3$  \\  
18140441-0822089     & 202221  &  CGMW3-01931            &  21.20 &  4.39 &  7.00 & 11.05 & 1.70 & 0.49 &     3.4  &            \\  
18472685-2331404     & 208511  &  CGMW4-03110            &  11.35 & -9.66 &  2.24 & 11.44 & 1.06 & 0.29 &     4.1  &            \\  
18494179-2058461     &         &  2MFGC 14602            &  13.91 & -9.03 &  1.32 & 10.91 & 0.96 & 0.08 &     3.8  &            \\  
19113405-0526417     & 203653  &  CGMW3-03439            &  30.36 & -6.95 &  2.40 & 10.02 & 1.22 & 0.35 &     4.6  &            \\  
19321420+3629565     &         &                         &  69.84 &  8.31 &  0.54 & 13.84 & 1.45 & 0.67 &     2.7  &            \\  
20021959+4307366     & 3097214 &                         &  78.52 &  6.55 &  4.47 &  9.67 & 1.32 & 0.42 &     3.9  &  5003$^4$  \\  
20491597+5119089     &         &                         &  89.84 &  4.73 &  6.29 &  7.39 & 1.03 & 4.65 &     3.7  &            \\  
21512406+4603029     & 167543  &  ZOAG093.34-06.22       &  93.35 & -6.22 &  1.57 & 10.34 & 1.19 & 0.32 &     3.5  &            \\  
21541934+5742569     &         &                         & 101.01 &  2.59 & 18.46 & 10.06 & 1.95 & 0.87 &     4.1  &            \\  
22064920+5902346     &         &                         & 103.12 &  2.66 & 10.23 &  9.82 & 1.62 & 0.55 &     5.2  &            \\  
22342113+5759454     &         &                         & 105.57 & -0.18 &  5.09 &  9.45 & 1.53 & 0.45 &     7.7  &            \\  
22571921+6155398     &         &                         & 109.94 &       & 17.02 & 10.37 & 1.86 & 0.88 &     4.6  &            \\  
23085314+6133564     &         &                         & 111.04 &  1.08 & 11.10 & 11.14 & 1.93 & 0.92 &     6.4  &            \\  
23271620+6900463     & 2726595 &                         & 115.48 &  7.37 &  4.06 & 10.06 & 1.37 & 0.45 &     6.5  &            \\  
23352762+6452140     &         &                         & 114.97 &  3.18 & 22.69 & 12.61 & 1.93 & 0.91 &     7.1  &            \\  
23471906+6027148     &         &                         & 115.10 & -1.44 & 20.58 &  9.03 & 1.98 & 0.84 &     2.0  &            \\  
\hline
\multicolumn{11}{l}{{\bf Notes}: Not listed are the 12 2MASX sources with extremely high, continuum perturbed rms noise levels (see Sect. 4). A $\star$ }  \\
\multicolumn{11}{l}{indicates sources searched for in the -275 to 11,225 \kms\ range, all others   were searched for in the 325 to 11,825 \kms\ range.}  \\
\multicolumn{11}{l}{References to optical redshifts: (1) Huchra et al. (1983), (2) Wegner et al. (2003), (3) West et al. (1981), (4) Tully (2002).}  \\
\end{tabular}
}
\end{table}

\newpage
\clearpage 

\begin{landscape}
\begin{table*}  
\caption{Arecibo: observational data}
\centering
{\footnotesize
\begin{tabular}{lrrlrlrrrrrrrrrrrrrrr}
\hline\hline             
No & 2MASX J          & PGC No  & Other Name        & $K_{20}$ & $J$-$K$& $H$-$K$& $r_{K20}$ & $b/a$ &$B_{T_C}$&$D_{25}$&$I_{HI}$&$V_{50}$&$W_{50}$&$W_{20}$&$rms$&$S_{max}$  \\
   & & & & mag & mag  &  mag & $''$ & & mag & $'$ & \JYKMS\ & \KMS\ & \KMS\ & \KMS\ & mJy & mJy   \\
(1)  &  (2) & (3) & (4) & (5) & (6) & (7) & (8) & (9) & (10) & (11) & (12) & (13) & (14) & (15) & (16) & (17)  \\
\hline
A01 & 04545536+3454433 &   143016 & CAP 0451+34          & 12.32 & 1.46 & 0.48 & 19.80 &  0.34 & 14.76 &  0.71 &  6.1$\pm$0.3 &  11433$\pm$4 & 434 & 469 &  1.94 &  26  \\
A02 & 05005348+3436115 &        * & *                    & 12.31 & 1.43 & 0.45 & 15.50 &  0.66 &     * &     * &  2.7$\pm$0.2 &  15352$\pm$2 & 250 & 266 &  1.51 &  18  \\
A03 & 05103476+2838002 &  1841450 & *                    & 13.31 & 1.37 & 0.44 &  6.10 &  0.80 & 15.69 &  0.32 &  1.3$\pm$0.2 & 17301$\pm$11 & 288 & 333 &  1.00 &   6  \\
A04 & 05112278+3621199 &        * & *                    & 13.87 & 1.43 & 0.71 &  8.70 &  0.39 &     * &    * & 0.56$\pm$0.13 &  7861$\pm$8  & 191 & 220 &  1.04 &   6  \\
A05 & 05165184+2919457 &        * & 2MFGC 4312           & 13.09 & 1.17 & 0.59 & 10.80 &  0.30 & 15.10 & 0.35 & 0.32$\pm$0.11 & 14414$\pm$10 & 128 & 155 & 1.03 &   4  \\
A06 & 05183772+3502015 &        * & *                    & 12.52 & 1.51 & 0.49 & 14.70 &  0.46 &     * &     * &  1.0$\pm$0.2 &  9008$\pm$3  & 338 & 350 &  1.18 &   7  \\
A07 & 05214377+1923370 &  1590726 & *                    & 13.78 & 0.96 & 0.24 & 10.40 &  0.46 & 15.68 &  0.62 &  2.3$\pm$0.1 &  4360$\pm$2  & 203 & 217 &  1.08 &  15  \\
A08 & 05315137+1517480 &  3097140 & *                    & 11.75 & 1.16 & 0.34 & 23.10 &  0.46 &     * &     * &  2.6$\pm$0.1 &  5812$\pm$2  & 180 & 206 &  1.14 &  20  \\
A09 & 05384324+1616012 &        * & *                    & 12.35 & 1.03 & 0.22 & 16.00 &  0.46 &     * &     * &  1.5$\pm$0.2 &  5284$\pm$2  & 262 & 273 &  1.27 &   9  \\
A10 & 05393262+3031377 &        * & *                    & 13.01 & 1.48 & 0.64 &  9.50 &  0.66 &     * &     * &  1.7$\pm$0.1 &  9384$\pm$2  & 114 & 137 &  1.01 &  20  \\
A11 & 05420143+1856156 &        * & *                    & 13.50 & 1.28 & 0.44 &  6.30 &  0.80 &     * &     * &  1.3$\pm$0.2 &  7603$\pm$64 & 216 & 414 &  1.08 &   5  \\
A12 & 05422061+2448359 &        * & IRAS 05393+2447      & 13.38 & 1.87 & 0.88 & 12.20 &  0.39 &     * &    * & 0.79$\pm$0.10 &  529$\pm$18  &  95 & 179 &  0.88 &   6  \\
A13 & 05453428+1307035 &  1423303 & 2MFGC 4668           & 12.60 & 1.11 & 0.32 & 20.60 &  0.30 & 15.28 &  0.51 &  3.5$\pm$0.1 &  7325$\pm$4  & 279 & 331 &  0.86 &  15  \\
A14 & 05490625+1904314 &   136129 & ZOAG 188.86-04.41    & 12.09 & 1.19 & 0.35 & 21.60 &  0.30 &     * &  0.40 &  1.5$\pm$0.1 &  5782$\pm$3  & 312 & 327 &  0.85 &   7  \\
A15 & 05594377+3224327 &        * & 2MFGC 4834           & 11.93 & 1.49 & 0.43 & 30.40 &  0.14 &     * &     * &  1.7$\pm$0.1 &  7765$\pm$2  & 398 & 412 &  0.81 &   8  \\
A16 & 05595654+2850437 &  1847776 & *                    & 12.94 & 1.25 & 0.38 & 10.60 &  0.56 & 15.97 &  0.36 &  0.7$\pm$0.1 &  9406$\pm$29 & 259 & 365 &  0.88 &   5  \\
A17 & 06001509+1537007 &   136218 & ZOAG 193.20-03.85    & 12.32 & 1.11 & 0.31 & 19.40 &  0.32 &     * &  0.98 & 0.56$\pm$0.09 & 5459$\pm$5  & 206 & 272 &  0.64 &  14  \\
A18 & 06023193+2833375 &   136013 & ZOAG 182.20+03.01    & 13.01 & 1.29 & 0.42 & 13.10 &  0.32 & 16.25 &  0.31 &  1.4$\pm$0.1 &  8584$\pm$11 & 101 & 166 &  1.01 &   9  \\
A19 & 06062918+2623371 &   136075 & ZOAG 184.52+02.71    & 14.20 & 1.07 & 0.27 & 10.30 &  0.24 & 13.44 &  1.03 &  4.0$\pm$0.1 &  2715$\pm$1  & 205 & 229 &  1.04 &  26  \\
A20 & 06103483+2602595 &   136084 & ZOAG 185.27+03.35    & 12.27 & 1.29 & 0.44 & 18.50 &  0.30 & 15.95 &  0.40 &  1.3$\pm$0.2 &  9398$\pm$11 & 323 & 361 &  1.19 &   6  \\
A21 & 06181671+2911295 &   143067 & CAP 0615+29a         & 12.20 & 1.17 & 0.41 & 15.50 &  0.74 & 15.21 &  0.48 &  1.2$\pm$0.2 & 10538$\pm$8  & 293 & 323 &  1.19 &   7  \\
A22 & 06242643+2213057 &        * & *                    & 13.40 & 1.36 & 0.56 &  9.40 &  0.68 &     * &    * & 0.75$\pm$0.07 &  1404$\pm16$ &  50 & 101 &  0.86 &   4  \\
A23 & 06261975+2320174 &   143077 & *                    & 11.89 & 1.11 & 0.27 & 20.10 &  0.44 & 14.32 &  0.81 &  2.4$\pm$0.1 &  4547$\pm$2  & 230 & 251 &  0.75 &  14  \\
A24 & 06301575+1646422 &   136308 & ZOAG 195.62+03.04    & 11.67 & 1.10 & 0.32 & 20.20 &  0.42 & 14.71 &  0.59 & 10.5$\pm$0.1 &  2525$\pm$1  & 267 & 285 &  0.81 &  55  \\
A25 & 06385063+1155111 &   136449 & ZOAG 200.89+02.66    & 12.77 & 1.36 & 0.46 & 12.60 &  0.40 & 15.37 &  0.32 &  2.1$\pm$0.2 &  7536$\pm$3  & 294 & 312 &  1.15 &  11  \\
A26 & 07032646+1249080 &  1416812 & *                    & 12.53 & 1.00 & 0.34 & 14.80 &  0.40 & 15.67 &  0.63 &  1.1         &  7931$\pm$3  & 304 & 315 &  0.91 &   5  \\
A27 & 07055211+1451243 &  1468922 & *                    & 13.04 & 1.05 & 0.32 & 10.80 &  0.52 & 16.08 &  0.59 &  2.6$\pm$0.2 &  4623$\pm$2  & 222 & 241 &  0.91 &  17  \\
A28 & 20114449+1526113 &        * & *                    & 12.30 & 1.04 & 0.34 & 15.40 &  0.44 &     * &     * &  1.2$\pm$0.1 &  6352$\pm$8  & 276 & 316 &  0.66 &   5  \\
A29 & 20312562+2258155 &  1679168 & 2MFGC 15573          & 13.40 & 1.46 & 0.46 & 11.80 &  0.30 & 17.53 &  0.52 &  1.4$\pm$0.1 &  8237$\pm$3  & 289 & 307 &  0.88 &   7  \\
\hline
\end{tabular}
}
\vfill
\end{table*}
\end{landscape}


\newpage
\clearpage 

\begin{landscape}
\begin{table*}  
\caption{Arecibo: derived data}
\centering
{\footnotesize
\begin{tabular}{crrrrlrrrrrrrrrrrrrrrr}
\hline\hline             
No. &   $l$  &    $b$ & $A_B$ &  $D$  &$k_{J-K}$& $R_{K20}$ &log($M_{HI}$)&log($L_K$)&log$({M_{HI}\over L_K})$&
log($L_B$)& {\bf log$({M_{HI}\over L_B})$} &$v_{rot}$&log($M_{dyn}$)&log$({M_{baryon}\over M_{dyn}})$   \\
    &   deg  &    deg & mag & Mpc   & mag  & kpc  & \Msun\ & \LsunK\ &   & \LsunB\ &       & \KMS\ & \Msun\ &  \\
(1)  &  (2)  & (3)    & (4) & (5)   & (6)  & (7)  & (8)   & (9)   & (10)  & (11)  & {\bf (12)}  & (13)  & (14)  & (15) \\
\hline
A01 & 168.95 & -5.41 & 3.97 & 163.8 & 0.10 & 15.7 & 10.58 & 10.96 & -0.37 & 10.72 & -0.14 & 230.7 & 11.28 & -0.18 \\  
A02 & 169.95 & -4.64 & 3.47 & 220.0 & 0.13 & 16.5 & 10.49 & 11.20 & -0.72 &     * &     * & 166.4 & 11.02 &  0.21 \\  
A03 & 176.00 & -6.56 & 3.34 & 247.8 & 0.14 &  7.3 & 10.28 & 10.90 & -0.62 & 10.70 & -0.42 & 240.0 & 10.99 & -0.03 \\  
A04 & 169.84 & -1.88 & 4.40 & 110.8 & 0.06 &  4.7 &  9.21 & 10.02 & -0.80 &     * &     * & 103.7 & 10.06 & -0.03 \\  
A05 & 176.23 & -5.04 & 2.65 & 206.5 & 0.12 & 10.8 &  9.51 & 10.81 & -1.30 & 10.78 & -1.27 &  67.1 & 10.05 &  0.70 \\  
A06 & 171.77 & -1.45 & 4.38 & 127.9 & 0.07 &  9.1 &  9.60 & 10.68 & -1.08 &     * &     * & 190.3 & 10.88 & -0.23 \\  
A07 & 185.14 & -9.72 & 1.86 &  58.6 & 0.03 &  3.0 &  9.26 &  9.41 & -0.15 &  9.46 & -0.20 & 114.3 &  9.95 & -0.29 \\ 
A08 & 189.96 & -9.89 & 2.87 &  79.5 & 0.05 &  8.9 &  9.60 & 10.52 & -0.93 &     * &     * & 101.4 & 10.32 &  0.19 \\
A09 & 190.00 & -7.98 & 2.16 &  71.5 & 0.04 &  5.5 &  9.25 & 10.17 & -0.92 &     * &     * & 147.5 & 10.44 & -0.28 \\  
A10 & 177.95 & -0.28 & 5.24 & 133.3 & 0.08 &  6.1 &  9.86 & 10.55 & -0.69 &     * &     * &  75.9 &  9.91 &  0.68 \\  
A11 & 188.12 & -5.92 & 1.87 & 106.4 & 0.06 &  3.3 &  9.53 & 10.04 & -0.51 &     * &     * & 180.0 & 10.38 & -0.25 \\ 
A12 & 183.14 & -2.78 & 4.42 &   8.8 & 0.01 &  0.5 &  7.16 &  8.01 & -0.85 &     * &     * &  51.6 &  8.50 & -0.49 \\  
A13 & 193.59 & -8.18 & 2.28 & 102.3 & 0.06 & 10.2 &  9.94 & 10.38 & -0.44 & 10.10 & -0.16 & 146.2 & 10.70 & -0.20 \\  
A14 & 188.86 & -4.41 & 3.19 &  78.6 & 0.05 &  8.2 &  9.34 & 10.39 & -1.05 &     * &     * & 163.5 & 10.70 & -0.34 \\
A15 & 178.55 &  4.38 & 3.06 & 108.7 & 0.06 & 16.0 &  9.69 & 10.73 & -1.04 &     * &     * & 201.0 & 11.17 & -0.47 \\ 
A16 & 181.67 &  2.65 & 2.10 & 133.4 & 0.08 &  6.9 &  9.49 & 10.47 & -0.98 & 10.05 & -0.56 & 156.3 & 10.59 & -0.14 \\  
A17 & 193.20 & -3.85 & 2.38 &  73.7 & 0.04 &  6.9 &  8.86 & 10.21 & -1.36 &     * &     * & 108.7 & 10.27 & -0.12 \\  
A18 & 182.20 &  3.01 & 1.80 & 121.0 & 0.07 &  7.7 &  9.68 & 10.35 & -0.66 &  9.86 & -0.18 &  53.3 &  9.70 &  0.69 \\  
A19 & 184.52 &  2.71 & 2.15 &  35.7 & 0.02 &  1.8 &  9.08 &  8.82 &  0.26 &  9.92 & -0.84 & 105.6 &  9.66 & -0.31 \\  
A20 & 185.27 &  3.35 & 2.44 & 133.2 & 0.08 & 11.9 &  9.75 & 10.75 & -1.00 & 10.06 & -0.31 & 169.3 & 10.90 & -0.17 \\ 
A21 & 183.30 &  6.34 & 2.49 & 150.0 & 0.09 & 11.3 &  9.81 & 10.88 & -1.07 & 10.46 & -0.65 & 217.8 & 11.09 & -0.24 \\  
A22 & 190.15 &  4.34 & 2.63 &  19.1 & 0.01 &  0.9 &  7.81 &  8.62 & -0.81 &     * &     * &  34.1 &  8.37 &  0.26 \\  
A23 & 189.35 &  5.24 & 1.82 &  60.0 & 0.04 &  5.8 &  9.32 & 10.19 & -0.87 & 10.02 & -0.70 & 128.1 & 10.34 & -0.15 \\  
A24 & 195.62 &  3.04 & 2.39 &  33.1 & 0.02 &  3.2 &  9.43 &  9.78 & -0.34 &  9.35 &  0.08 & 147.1 & 10.21 & -0.27 \\  
A25 & 200.89 &  2.66 & 2.58 & 104.8 & 0.06 &  6.4 &  9.74 & 10.34 & -0.60 & 10.09 & -0.35 & 160.4 & 10.58 & -0.18 \\  
A26 & 202.79 &  8.43 & 0.34 & 110.7 & 0.06 &  7.9 &  9.49 & 10.41 & -0.93 & 10.01 & -0.52 & 165.8 & 10.70 & -0.30 \\  
A27 & 201.19 &  9.85 & 0.39 &  61.2 & 0.04 &  3.2 &  9.36 &  9.69 & -0.34 &  9.33 &  0.03 & 129.9 & 10.10 & -0.24 \\  
A28 &  56.12 & -9.97 & 0.96 &  90.0 & 0.05 &  6.7 &  9.36 & 10.34 & -0.98 &     * &     * & 153.7 & 10.56 & -0.24 \\  
A29 &  65.10 & -9.69 & 0.88 & 118.2 & 0.07 &  6.8 &  9.66 & 10.14 & -0.47 &  9.33 &  0.33 & 151.5 & 10.55 & -0.31 \\  

\hline
\end{tabular}
}
\vfill
\end{table*}
\end{landscape}

\newpage
\clearpage

\begin{table}  
\caption{Arecibo: non-detections}
 \centering
{\footnotesize
\begin{tabular}{lrlrrrrrrrrl}
\hline\hline
2MASX J          &  PGC       & others               & $l$   &   $b$ & $A_B$ & $K_{20}$ & $J$-$K$ & $H$-$K$ & rms   & rms   & V$_{opt}$ \\ 
                 &            &                      & (deg) & (deg) & (mag) & (mag)    & (mag)   & (mag)   & (mJy) & (mJy) & (\kms)    \\
\hline
04595036+2812274   &         &                       & -8.71  & 0.72  & 2,98 & 13.08  & 0.90  & 0.69  &  0.90   &         &        \\ 
05191235+2817462   &         &                       & -5.21  & 0.65  & 2,69 & 14.13  & 0.97  & 0.34  &  0.95   &  1.03   &        \\ 
05210718+2956105   &         &                       & -3.93  & 0.60  & 2,48 & 13.29  & 1.10  & 0.59  &  1.01   &  1.15   &        \\ 
05254740+2440008   &         &                       & -6.02  & 1.17  & 4,84 & 13.19  & 1.00  & 0.36  &  1.99   &         &        \\ 
05264832+2905181   &         &                       & -3.38  & 0.54  & 2,24 & 14.20  & 0.93  & 0.20  &  1.11   &  0.95   &        \\ 
05332198+1936518   &         &                       & -7.30  & 0.42  & 1,74 & 13.08  & 1.10  & 0.71  &  1.69   &         &        \\ 
05344787+1805357   &         &                       & -7.82  & 0.50  & 2,07 & 13.08  & 0.90  & 0.69  &  1.02   &  1.58   &        \\ 
05355814+2034267   & 1630996 &                       & -6.27  & 0.45  & 1,86 & 13.38  & 0.71  & 0.57  &  3.34   &         &        \\ 
05374258+2856044   &         &                       & -1.47  & 1.54  & 6,38 & 12.70  & 1.32  & 0.62  &  0.97   &         &        \\ 
05402712+1611389   & 1503020 &                       & -7.66  & 0.44  & 1,82 & 13.07  & 0.62  & 0.83  &  0.74   &  1.03   &        \\ 
05425096+2856282   &         &                       & -0.51  & 2.03  & 8,40 & 13.99  & 1.40  & 0.17  &  1.00   &         &        \\ 
05430547+2554272   &         &                       & -2.06  & 1.26  & 5,22 & 13.54  & 1.03  & 0.44  &  0.95   &         &        \\ 
05431284+2852070   &         &                       & -0.48  & 1.59  & 6,58 & 12.86  & 1.29  & 0.67  &  1.01   &         &        \\ 
05432537+2831300   &         &                       & -0.63  & 1.40  & 5,80 & 14.08  & 0.49  & 0.38  &  0.95   &         &        \\ 
05433224+2644020   &         &                       & -1.54  & 1.52  & 6,29 & 13.01  & 1.19  & 0.66  &  0.97   &         &        \\ 
05442930+3340232   &         &  ZOAG 175.83+02.26    &  2.27  & 0.75  & 3,11 & 13.45  & 0.82  & 0.35  &  1.07   &         &        \\ 
05443397+1537361   & 1489395 &  2MFGC 4651           & -7.11  & 0.35  & 1,45 & 12.09  & 0.75  & 0.36  &  0.96   &  1.14   &        \\ 
05454063+2335051   &         &                       & -2.77  & 1.04  & 4,31 & 13.27  & 1.07  & 0.50  &  1.24   &         &        \\ 
05522831+2722156   &         &                       &  0.49  & 1.07  & 4,43 & 12.91  & 0.99  & 0.55  &  1.05   &         &        \\ 
05542241+1759203   & 136165  &  ZOAG 190.44-03.88    & -3.89  & 0.76  & 3,15 & 12.99  & 0.97  & 0.34  &  0.88   &         &        \\ 
05542430+1211464   &         &                       & -6.77  & 0.35  & 1,45 & 13.25  & 1.01  & 0.58  &  0.86   &         &        \\ 
05562009+2913201   & 135940  &  ZOAG 180.95+02.16    &  2.16  & 0.60  & 2,48 & 13.33  & 0.84  & 0.54  &  0.95   &         &        \\ 
05574013+1400482   &         &  2MFGC 4810           & -5.18  & 0.37  & 1,53 & 12.91  & 0.97  & 0.57  &  0.86   &         &        \\ 
05575057+2243246   &         &                       & -0.81  & 1.30  & 5,38 & 13.37  & 1.14  & 0.36  &  0.84   &         &        \\ 
05575200+1829246   &         &                       & -2.92  & 1.33  & 5,51 & 12.87  & 0.99  & 0.52  &  0.78   &         &        \\ 
05580146+2215426   &         &                       & -1.00  & 1.12  & 4,64 & 13.18  & 1.13  & 0.66  &  0.82   &         &        \\ 
05584853+1422008   &         &                       & -4.77  & 0.37  & 1,53 & 13.05  & 0.80  & 0.51  &  1.07   &         &        \\ 
05594673+2515307   &         &                       &  0.84  & 1.13  & 4,68 & 12.40  & 1.06  & 0.33  &  0.96   &         &        \\ 
06010650+2234214   &         &  2MFGC 4853           & -0.23  & 1.27  & 5,26 & 13.04  & 1.25  & 0.52  &  0.98   &         &        \\ 
06020649+2955522   & 135941  &                       &  3.60  & 0.42  & 1,74 & 12.77  & 0.75  & 0.46  &  1.13   &         &        \\ 
06024795+2928434   & 135968  &  ZOAG 181.43+03.51    &  3.51  & 0.51  & 2,11 & 12.64  & 0.82  & 0.44  &  1.40   &         &        \\ 
06024888+2940221   & 1874057 &  ZOAG 181.26+03.60    &  3.61  & 0.49  & 2,03 & 13.21  & 0.83  & 0.45  &  0.88   &         &        \\ 
06151022+2522405   & 136100  &  ZOAG 186.35+03.94    &  3.94  & 0.80  & 3,31 &  9.96  & 0.88  & 0.35  &  1.09   &         &        \\ 
06215413+1945476   &         &  2MFGC 5125           &  2.67  & 0.72  & 2,98 & 13.18  & 1.09  & 0.47  &  0.81   &         &        \\ 
06274322+2653401   & 1791580 &                       &  7.14  & 0.35  & 1,45 & 13.28  & 0.71  & 0.45  &  0.87   &         &        \\ 
06284058+1655177   & 97169   &  ZOAG 195.31+02.77    &  2.77  & 0.55  & 2,28 & 12.23  & 0.77  & 0.44  &  0.88   &         & 4304$^5$ \\ 
06284437+2430089   & 1708860 &                       &  6.26  & 0.35  & 1,45 & 13.54  & 0.78  & 0.64  &  0.92   &         &        \\ 
06312775+2447297   & 1716505 &                       &  6.94  & 0.34  & 1,41 & 13.52  & 0.86  & 0.25  &  1.06   &         &        \\ 
06313318+2416319   & 1703633 &                       &  6.73  & 0.32  & 1,32 & 13.36  & 0.79  & 0.51  &  0.96   &         &        \\ 
06313599+2551219   & 1752511 &                       &  7.45  & 0.31  & 1,28 & 13.00  & 0.86  & 0.55  &  0.82   &         &        \\ 
06324150+1645169   &         &                       &  3.54  & 0.73  & 3,02 & 12.72  & 0.81  & 0.53  &  0.75   &         &        \\ 
06345218+1657036   &         &                       &  4.10  & 0.65  & 2,69 & 13.03  & 0.86  & 0.50  &  0.79   &         &        \\ 
06360012+2359092   &         &                       &  7.51  & 0.16  & 0,66 & 13.29  & 0.69  & 0.51  &  0.83   &         &        \\ 
06361416+1717163   &         &                       &  4.54  & 0.53  & 2,19 & 13.48  & 1.04  & 0.37  &  0.83   &         &        \\ 
06380355+2447381   &         &                       &  8.28  & 0.17  & 0,70 & 13.47  & 0.91  & 0.51  &  1.01   &         &        \\ 
06383300+2354002   &         &                       &  7.99  & 0.16  & 0,66 & 12.90  & 0.84  & 0.39  &  0.90   &         &        \\ 
06383995+2345504   &         &  2MFGC 5306           &  7.95  & 0.18  & 0,75 & 12.99  & 0.91  & 0.55  &  1.02   &         &        \\ 
06400101+1851424   &         &                       &  6.05  & 0.33  & 1,37 & 13.46  & 0.70  & 0.68  &  0.72   &  0.67   &        \\ 
06443767+1339026   &         &  ZOAG 199.99+04.70    &  4.70  & 0.33  & 1,37 & 12.85  & 0.86  & 0.58  &  1.10   &  0.97   &        \\ 
06444477+1537015   &         &                       &  5.61  & 0.20  & 0,83 & 13.31  & 0.70  & 0.45  &  0.93   &         &        \\ 
06594659+1406580   &         &  2MFGC 5562           &  8.20  & 0.09  & 0,37 & 12.85  & 0.76  & 0.24  &  0.86   &         &        \\ 
20021227+1236353   &         &                       & -9.46  & 0.16  & 0,66 & 13.45  & 0.70  & 0.45  &  0.89   &         &        \\ 
\hline
\multicolumn{12}{l}{{\bf Notes}: rms noise levels are for searches in the range of -500 to 11,000 \kms\ (Col. 4) and 9,500 to 21,000 \kms\ (Col. 5).}  \\
\multicolumn{12}{l}{References to optical redshifts: (5) Takata et al. (1994). }  \\
\end{tabular}
}
\end{table}

\newpage
\clearpage

\begin{table}  
\caption{Comparison with published \HI\ data}
 \centering
 {\footnotesize
\begin{tabular}{lllrll}
\hline\hline
No       & \IHI\        &  \VHI\      & $W_{50}$ & telescope & reference  \\
         & Jy km/s      & km/s       & km/s  &              &            \\
\hline
N06        &  1.1$\pm$0.3 &  2435$\pm$5 &  80 & \nan\   & present study  \\
           &  1.7         &  2458       &  72 & \nan\   & Theureau et al. 1998  \\
N10        &  8.2$\pm$0.6 &  3919$\pm$3 & 357 & \nan\   & present study   \\
           &  7.1         &  3913       & 356 & \nan\   & Paturel et al. (2003b)  \\
N11        & 15.1$\pm$0.2 &  2863$\pm$1 &  47 & \nan\   & present study  \\
           & 11.9         &  2861       &  37 & Parkes  & Henning et al. (2000)  \\ 
N12        &  5.8$\pm$0.5 &  6575$\pm$3 & 463 & \nan\   & present study  \\
           & ---          &  6550       & 480 & Effelsberg  & from Huchtmeier et al. (2005)  \\
$[$N15$]$  &  2.2         &  1302       &  73 & \nan\   & present study  \\
           &  2.5         &  1313       &  89 & \nan\   & Chamaraux et al. (1999)  \\
ESO 371-27 &  4.9         &  1317       &     & Parkes  & Doyle et al. (2005)  \\ 
N17        & 31.3$\pm$0.1 &  1767$\pm$1 & 143 & \nan\   & present study  \\
           & 35.3         &             & 137 & Parkes  & Ryan-Weber et al. (2002)  \\ 
N18        &  9.2$\pm$0.3 &  1786$\pm$1 & 313 & \nan\   & present study  \\
           &  12.6        &  1788       & 317 & Parkes  & Meyer et al. (2004)  \\   
A01        &  6.1$\pm$0.3 & 11433$\pm$4 & 434 & Arecibo & present study   \\
           &  4.75        &   11425     & 447 & Arecibo & Pantoja et al. (1997)  \\
A19        &  4.0$\pm$0.1 &  2715$\pm$1 & 205 & Arecibo & present study   \\
           & 5.35         &  2717       & 208 & Arecibo & Pantoja et al. (1997)  \\
A21        &  1.2$\pm$0.2 & 10538$\pm$8 & 293 & Arecibo & present study   \\
           &  1.52        & 10525       & 287 & Arecibo & Pantoja et al. (1997)  \\
A23        &  2.4$\pm$0.1 &  4547$\pm$2 & 230 & Arecibo & present study   \\
           & 2.16         &  4545       & 230 & Arecibo & Pantoja et al. (1997)  \\
A24        & 10.5$\pm$0.1 &  2525$\pm$1 & 267 & Arecibo & present study   \\
           & 30.3         &  2532       & 258 & Parkes  & Donley et al. (2005)  \\
\hline
\multicolumn{6}{l}{{\bf Notes}: no. N15: galaxy ESO 371-27 was detected in the beam centred on the undetected }  \\
\multicolumn{6}{l}{target, ESO 371-26 (see Sect. 4.1). }  \\
\end{tabular}
}
\end{table}
\twocolumn

\end{document}